\documentclass[pre,twocolumn,aps]{revtex4-2}
\usepackage{amsmath}
\usepackage{amssymb}
\usepackage{amsthm}
\usepackage{graphicx}
\usepackage{bm}
\usepackage{physics}
\usepackage{mathtools}
\usepackage{color}
\usepackage[caption=false]{subfig}

\usepackage{afterpage}

\newtheorem*{theorem*}{Theorem}

\usepackage{hyperref}
\definecolor{myblue}{RGB}{0,0,255}
\hypersetup{
    colorlinks,
    citecolor=myblue,
    linkcolor=myblue,
    urlcolor=myblue
}

\usepackage{subfig}
\usepackage{stmaryrd}

\usepackage{booktabs}

\begin{document}

\title{Quantum-computer-based verification of quantum thermodynamic uncertainty relation}
\author{Nobumasa Ishida}
\email{ishida@biom.t.u-tokyo.ac.jp}
\affiliation{Department of Information and Communication Engineering, Graduate School of Information Science and Technology, The University of Tokyo, Tokyo 113-8656, Japan}
\author{Yoshihiko Hasegawa}
\email{hasegawa@biom.t.u-tokyo.ac.jp}
\affiliation{Department of Information and Communication Engineering, Graduate School of Information Science and Technology, The University of Tokyo, Tokyo 113-8656, Japan}

\begin{abstract}
Quantum thermodynamic uncertainty relations establish fundamental trade-offs between the precision achievable in quantum systems and associated thermodynamic quantities such as entropy production or dynamical activity. While foundational, empirical demonstrations have thus far been confined to specific cases, either assuming time-reversal symmetry or involving particular measurement types, leaving the verification of their universal validity unrealized. This work leverages a quantum computer to present the first empirical verification of a general quantum thermodynamic uncertainty relation, valid for arbitrary dynamics and observables. We theoretically derive the relation, identifying survival activity as the pivotal thermodynamic quantity governing the precision bound. The verification is demonstrated on IBM's cloud-based quantum processor, which is treated as a real thermodynamic system. To achieve accurate results despite substantial device errors, we introduce a generic protocol for measuring survival activity and employ circuit reduction techniques based on the relation's properties. This strategy allows us to empirically measure survival activity for the first time and confirm the derived relation. Furthermore, the quantum computer's versatility enables the implementation of optimal observables, leading to the saturation of the relation and demonstrating the sharpness of our bound on a physical device. The method's broad applicability is further illustrated by verifying the trade-off for quantum time correlators. Our findings establish quantum computers as effective platforms for investigating fundamental thermodynamic trade-off relations.
\end{abstract}

\maketitle

\section{Introduction}
Understanding the fundamental limits of quantum system performance is essential to advance quantum information processing. Quantum thermodynamics has recently revealed these limits via trade-off relations \cite{Huber2015-hh,Campbell2017-wp,Vitagliano2018-ig,Chitambar2019-rx,Funo2019-cm,Pearson2021-ep,Hasegawa2023-de}, which offer general predictions for the attainable performance under thermodynamic constraints. Among the most significant are quantum thermodynamic uncertainty relations (QTURs), bounding the precision of observables in terms of thermodynamic quantities such as entropy production and dynamical activity \cite{Erker2017-hb,Brandner2018-cq,Carollo2019-lk,Timpanaro2019-eh,Saryal2019-tg,Guarnieri2019-sy,Liu2019-br,Hasegawa2020-rh,Miller2021-zn,Liu2021-um,Kalaee2021-oi,Hasegawa2021-aq,Hasegawa2022-xg,Van_Vu2022-hm,Monnai2022-uc,Hasegawa2023-de}. The original thermodynamic uncertainty relation was formulated for classical systems in steady states \cite{Barato2015-fy,Gingrich2016-vf}:
\begin{align}
    \frac{{\rm Var}[\mathcal{F}_{\rm cl}]}{\expval{\mathcal{F}_{\rm cl}}^2}\geq \frac{2}{\Sigma},\label{eq:classical-TUR}
\end{align}
where $\mathcal{F}_{\rm cl}$ is a classical current-like observable and $\Sigma$ is the entropy production. This inequality implies that reducing relative fluctuations, quantified by the ratio of the variance to the square of the mean $\langle\mathcal{F}_{\rm cl}\rangle$, requires increased dissipation. This classical bound has since been extended to a broad class of systems \cite{Horowitz2019-pl,Kwon2024-up}. Building upon this development, recent studies have focused on quantum regimes, where quantum features such as coherence \cite{Kalaee2021-oi,Prech2025-yr} may influence precision. QTURs are thus emerging as universal performance bounds for quantum processes, from the accuracy of quantum clocks \cite{Erker2017-hb,Hasegawa2022-xg,He2023-zn,Meier2023-xv} to the reliability of quantum heat engines \cite{Miller2021-zn}, with potential to generalize beyond precision trade-offs \cite{Pietzonka2018-pe,Van_Tuan2022-nm,Falasco2022-ku,Kwon2024-up}. 

A central challenge in developing a fundamental QTUR, from the standpoint of predictive ability, is to simultaneously satisfy three criteria: (i) \textit{generality}, (ii) \textit{sharpness}, and (iii) \textit{empirical accessibility}. Generality requires that the relation holds for arbitrary quantum dynamics and observables, irrespective of system-specific details. Sharpness demands the bound is tight---ideally being saturable---for certain practical conditions, thereby highlighting when precision is maximized \cite{Dechant2021-di,Koyuk2022-st,Dieball2023-ll}. However, these two properties often conflict with the third: accessibility. Theoretical efforts to derive general bounds have frequently introduced cost functions that are difficult, if not impossible, to measure in practice. For instance, a QTUR applicable to some observables on general quantum channels, which are described by completely positive trace preserving (CPTP) maps, was recently derived \cite{Hasegawa2021-aq}. Yet its cost function, the ``survival activity,'' is defined via matrix inverse related to the quantum channel, posing an observation obstacle. Otherwise, maintaining conventional costs such as entropy production or dynamical activity results in the introduction of abstract correction terms that are less physically observable \cite{Koyuk2020-qm,Van_Vu2022-hm,Salazar2024-de,Kwon2024-up}.

Consequently, empirical verification of QTURs has been confined to specific scenarios: either time-reversal symmetric dynamics accompanied by observables with definite parity \cite{Friedman2020-fo,Pal2020-os}, or continuous measurement protocols \cite{He2023-zn}. Therefore, a general, tight, and empirically verifiable QTUR has remained elusive. This leaves key questions unanswered: What is the fundamental thermodynamic quantity that governs precision in any quantum system, and can this universal principle be empirically verified and proven effective on realistic devices?

In this work, we address these questions by demonstrating a general and saturable QTUR through quantum-computer-based verification. This bound is derived in the present study and applies to arbitrary observables under general quantum dynamics governed by CPTP maps, thereby extending the scope of prior QTURs. We identify the survival activity \cite{Hasegawa2021-aq} as the pivotal thermodynamic quantity that determines the precision limit. The relation is saturable with optimal observables, and crucially, we validate these features empirically.

We introduce quantum computers as a testbed for investigating thermodynamic trade-off relations. Our approach treats the quantum processor itself as a thermodynamic system, exploring empirical characteristics such as finite qubit temperatures and dissipative dynamics. This contrasts with previous works that focus on simulating idealized models \cite{Fauseweh2024-bj,Bauer2020-rv,Bauer2023-qf}. A distinctive advantage of quantum computers over conventional platforms for QTUR experiments is their ability to probe arbitrary observables via flexible measurement schemes. Specifically, we implement optimal observables that saturate the QTUR using collective measurements, thereby extracting the maximum precision from the device.

A major obstacle to quantum-computer-based verification stems from the limited capabilities of current hardware: short coherence times and large gate errors severely restrict the applicable circuit depth \cite{Preskill2018-ug,Zhou2020-qu,Cheng2023-gd}. To overcome these constraints, we develop a methodology for accurately measuring the physical quantities relevant to our QTUR. First, we propose a generic protocol to measure survival activity using a Neumann series expansion, which avoids intractable matrix inversion. Next, we introduce a circuit minimization technique that leverages the structure of the QTUR. This reduces the circuit depth by an order of magnitude. As we show, without such reduction, the accuracy needed for trade-off verification would be out of reach.

Using IBM's cloud-based quantum processor, we demonstrate (i) the first empirical identification of survival activity, (ii) validation of the general QTUR, and (iii) the first QTUR saturation using optimal observables on a physical device. These results highlight that our general bound is effective on current quantum devices. An analysis of empirical errors further suggests that finite qubit temperatures significantly influence precision limits, which indicates that our verification serves as a hardware benchmark and contributes to the growing interface between quantum computing and quantum thermodynamics \cite{Blok2025-fh}. To demonstrate the broader applicability of our QTUR and verification method, we also apply them to quantum time correlators, confirming the associated trade-off relations. Taken together, our findings establish quantum computers as viable platforms for probing thermodynamic trade-off relations.

This paper is organized as follows. In Section~\ref{section:theory}, we develop a general theory of the QTUR, clarifying both the applicability and tightness of the derived bound. Section~\ref{section:verification-methodology} outlines the quantum-computer-based verification scheme, including the use of quantum computers as thermodynamic testbeds, the measurement protocol for survival activity, and circuit optimization tailored to QTUR verification. Section~\ref{section:demonstrations} presents two demonstrations: one implementing the optimal observable to test the QTUR, and another investigating quantum time correlators. For each case, we explain the aim and design before analyzing the empirical results. Finally, Section~\ref{section:conclusion} offers discussion and concluding remarks, highlighting future directions and possible links to broader research in quantum thermodynamics and quantum computing.

\section{Theory}\label{section:theory}
\begin{figure*}[ht]
    \centering
    \includegraphics[width=\textwidth]{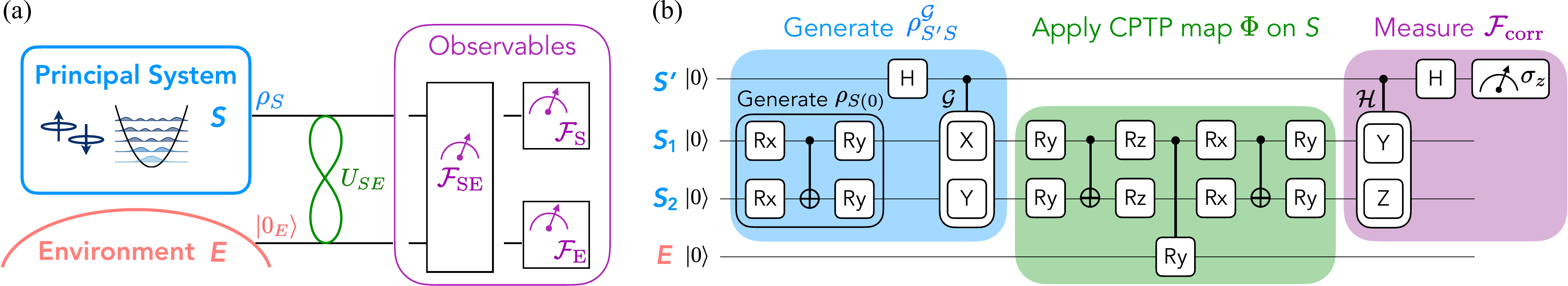}
    \caption[Illustration of the system and the main circuit]{Illustration of a quantum system comprising a principal system \textit{S} and its environment \textit{E}. (a) General description. The unitary evolution $U_{SE}$ of \textit{S}+\textit{E} induces a CPTP map $\Phi$ on \textit{S}. We focus on the expectation value and variance of generic observables $\mathcal{F}$, including local observables on \textit{S} or \textit{E}, for the final state of \textit{S}+\textit{E}. (b) Quantum circuit for measuring the quantum time correlator ${\rm Re}[R(T)]$, employed in the second demonstration of the general QTUR. Shaded areas correspond to the colored segments in (a). We take $\mathcal{G}=\sigma_x\otimes\sigma_y$ and $\mathcal{H}=\sigma_y\otimes\sigma_z$ as examples. $\mathcal{G}$ and $\mathcal{H}$ are randomly selected as tensor products of Pauli or identity operators, excluding $I\otimes I$. Rotation gates, including those for $\rho_S(0)$ preparation, are also randomly parameterized. The rotation angle of the controlled RY gate between $S$ and $E$ is $\pi\gamma$, where $\gamma$ is the interaction strength.}
    \label{fig:illustration}
\end{figure*}
\subsection{Setup}
We aim to derive a QTUR that serves as a bound to be tested, applying to arbitrary observables under any CPTP map. CPTP maps constitute the most general description of quantum processes, mapping one density matrix to another \cite{Nielsen2010-tb}. Consider a system \textit{S} and an environment \textit{E}, as shown in Fig.~\ref{fig:illustration}a. Initially, \textit{S} and \textit{E} are in states $\ket{\Psi_S(0)}$ and $\ket{0_E}$, respectively. The composite system \textit{S}+\textit{E} evolves under unitary operation $U_{SE}$ from $t=0$ to $T$, resulting in the final state $\ket{\Psi_{SE}(T)} = U_{SE}(\ket{\Psi_{S}(0)} \otimes \ket{0_E})$. Focusing on the environmental changes, we can also express $\ket{\Psi_{SE}(T)}$ as $\sum_m V_m\ket{\Psi_{S}(0)}\otimes\ket{m_E}$ with Kraus operators $V_m = \bra{m_E}U_{SE}\ket{0_E}$, where $\{\ket{m_E}\}$ is an orthonormal basis for \textit{E}. Specifically, $V_0$ is associated with the state of $S$ without any jumps in $E$. Then the CPTP map $\Phi$ on the reduced density matrix of $S$ is given by $\rho_{S}(T)=\Phi(\rho_{S}(0))=\sum_m V_m \rho_{S}(0) V_m^\dagger$.

While we assume $\rho_S(0)$ is pure for simplicity, the following theoretical results remain valid even if $\rho_S(0)$ is mixed, as shown in Appendix~\ref{section:appendix:derivation-of-generalQTUR}. Furthermore, in Appendix~\ref{section:appendix:generalization-to-thermal-environments}, we show that the theory can be extended to thermal environments. Nonetheless, our later demonstrations confirm that the QTUR with pure-state environments effectively describes physical qubits in quantum processors.

\subsection{General QTUR}
We derive the following QTUR bounding the precision of an arbitrary Hermitian observable $\mathcal{F}$ on $S$ and $E$:
\begin{equation}
    \frac{\operatorname{Var}[\mathcal{F}]}{(\expval{\mathcal{F}}-\mathcal{C}_\mathcal{F})^2}\geq\frac{1}{\mathcal{A}},\label{eq:GeneralTUR}
\end{equation}
which we refer to as the general QTUR, and its derivation is provided in Section~\ref{section:brief-derivation-generalQTUR}. Here, $\expval{\mathcal{F}} = \langle \Psi_{SE}(T) | \mathcal{F}| \Psi_{SE}(T) \rangle$ is the expectation value of $\mathcal{F}$ for the final state and $\text{Var}[\mathcal{F}] = \langle \mathcal{F}^2 \rangle - \langle \mathcal{F} \rangle^2$ is its variance. The crucial quantity $\mathcal{A}$ is the survival activity \cite{Hasegawa2021-aq}, defined as
\begin{align}
    \mathcal{A}=\expval*{(V_0^\dagger V_0)^{-1}}_{\rho_S(0)}-1,\label{eq:survial-activity}
\end{align}
with $\langle\cdot\rangle_\rho={\rm Tr}[\rho\cdot]$. As the expectation value $\langle V_0^\dagger V_0\rangle_{\rho_S(0)}$ represents the probability for the system $S$ to evolve under $V_0$ while the environment survives in its initial state $\ket{0_E}$, its reciprocal quantifies the environmental activity. $\mathcal{A}$ is associated with dynamical activity, which counts the number of jumps, for Markovian open quantum systems \cite{Hasegawa2021-aq}. Importantly, $\mathcal{A}$ depends only on the initial state $\rho_S(0)$ and the dynamics (via $V_0$), but is independent of the chosen observable $\mathcal{F}$. The correction term $\mathcal{C}_\mathcal{F}$ is an inherent-dynamics contribution specific to $\mathcal{F}$, detailed below.

The general QTUR [Eq.~\eqref{eq:GeneralTUR}] reveals that the precision of any observable $\mathcal{F}$ (quantified by its relative fluctuation) is constrained by the single, observable-independent thermodynamic quantity $\mathcal{A}$. The significance of the bound is underscored by three primary features:
\begin{enumerate}
    \item Applicability to any Hermitian observable on \textit{S}+\textit{E}: The observable $\mathcal{F}$ is generic. It can be global (i.e., an observable with an entangled eigenbasis across $S$+$E$), separable, independent ($\mathcal{F}_{S}\otimes \mathcal{F}_{E}$), or local ($\mathcal{F}_{S}\otimes I_{E}$ or $I_{S}\otimes \mathcal{F}_{E}$).
    \item Validity for arbitrary initial state of \textit{S} and any CPTP dynamics on \textit{S} provided that ${V_0}^{-1}$ exists.
    \item Sharpness: Equality holds for the optimal observable $\mathcal{L}$ [defined in Eq.~\eqref{eq:optimal-observable-L}].
\end{enumerate}

We comment on the generality of our bound. First, it applies to current-like observables, that is, the class of observables considered in the representative classical TURs in Eq.~\eqref{eq:classical-TUR} and Refs.~\cite{Hasegawa2019-se,Timpanaro2019-eh}. Moreover, the general QTUR bounds currents in quantum systems obtained via continuous measurement \cite{Landi2024-vn}, as discussed in Ref.~\cite{Hasegawa2021-aq}. Second, our result includes the existing QTUR for CPTP maps in Ref.~\cite{Hasegawa2021-aq}, which is restricted to environmental observables. We extend that bound to arbitrary observables on the joint system $S+E$, thereby enabling saturation. Despite certain differences, our approach aligns with the spirit of improving classical TURs by considering both system and environment \cite{Dechant2020-im,Dieball2023-ll}. Third, the assumption of a pure initial environment does not limit the applicability of our bound. As shown in Appendix~\ref{section:appendix:generalization-to-thermal-environments}, it can be extended to general thermal environments with minor modifications. Interestingly, though, our demonstrations show that the original bound [Eq.~\eqref{eq:GeneralTUR}] already describes qubits in quantum processors at low-but-finite temperatures.

The inherent-dynamics contribution $\mathcal{C}_\mathcal{F}$ is defined as the correlator
\begin{align}
    \mathcal{C}_\mathcal{F}={\rm Re}[\expval*{({V_0}^{-1}\otimes I_E) \mathcal{F}U_{SE}}_{\rho_{SE}(0)}],\label{eq:inherent-dynamics-contribution}
\end{align}
which captures the baseline part of $\expval{\mathcal{F}}$ arising solely from the no-jump dynamics governed by $V_0$. To illustrate, consider the local observable $\mathcal{F}_E = \sum_m m\ketbra{m_E}{m_E}$, which counts photons emitted from $S$ into $E$. Since no photons are emitted in the no-jump evolution, the expectation of $\mathcal{F}_E$ vanishes when E remains in $\ket{0_E}$, leading to $\mathcal{C}_\mathcal{F} = 0$ [Appendix~\ref{section:appendix:simplification-of-CF-for-Fs}]. In this case, the general QTUR reduces to the previously known bound \cite{Hasegawa2021-aq}. In contrast, $\mathcal{C}_\mathcal{F}$ becomes indispensable for observables that involve $S$. As a representative case, consider a system observable $\mathcal{F} = \mathcal{F}_S \otimes I_E$ under weak system-environment coupling. Here, $V_0$ approaches a unitary operator, $V_0^\dagger V_0 \approx I_S - \epsilon$ with a small-norm operator $\epsilon$. A perturbative analysis then yields $\mathcal{C_{F_S}}= \expval*{{V_0}^\dagger \mathcal{F}_S {V_0}}_{\rho_S(0)}+\mathcal{O}(\epsilon)$, where the leading term represents the expectation of $\mathcal{F}_S$ without any environmental jumps [Appendix~\ref{section:appendix:CF-weak-coupling}]. Generally, subtracting $\mathcal{C}_\mathcal{F}$ from $\expval{\mathcal{F}}$ in Eq.~\eqref{eq:GeneralTUR} isolates the contribution stemming from jump-induced processes: namely, those that genuinely reflect the interplay between $S$ and $E$. This separation is essential for the tightness and potential saturation of the bound, as $\mathcal{A}$ quantifies the fluctuations arising from these jump-mediated interactions. We provide concrete examples of $\mathcal{A}$ and $\mathcal{C}_\mathcal{F}$ in Appendix~\ref{section:appendix:derivation-U-SE-example}.

In the first demonstration described later [Section~\ref{section:demonstration-I}], we examine the empirical relevance of the general QTUR. Specifically, we confirm the measurability of $\mathcal{A}$ and $\mathcal{C}_\mathcal{F}$, the empirical validity of the bound, and the attainability of the equality condition on a real quantum device. These results suggest that the general QTUR may serve as a fundamental principle.

\subsection{Derivation}\label{section:brief-derivation-generalQTUR}
In this section, we outline the proof of the general QTUR [Eq.~\eqref{eq:GeneralTUR}]. For clarity, we assume an initial pure state for $S$. See Appendix~\ref{section:appendix:derivation-of-generalQTUR} for the full derivation, including the case of mixed initial states. While we employ a general CPTP map here to illustrate the overall picture, we present a concrete example with $U_{SE}$ used in our demonstrations in Appendix~\ref{section:appendix:derivation-U-SE-example}.

We use the quantum Cramér--Rao inequality from quantum estimation theory \cite{Helstrom1969-zv,Hotta2004-px,Paris2009-lx, Liu2019-iq}, which provides a lower bound on the variance of an estimator. We consider the parameter estimation problem for $\theta$, which virtually perturbs the CPTP map. The virtual perturbation is introduced via the following Kraus operators acting on \textit{S}, as in Ref.~\cite{Hasegawa2021-aq}:
\begin{align}
    V_m(\theta)=e^{\theta/2}V_m\;(m\geq1),\label{eq:end-matter-perturbation-Vm}\\
    V_0(\theta)=U_0\sqrt{I-e^{\theta}\sum_{m\geq1}^M V_m^\dagger V_m},\label{eq:end-matter-perturbation-V0}
\end{align}
where $U_0$ is a unitary operator in the polar decomposition $V_0=U_0\sqrt{V_0^\dagger V_0}$. The perturbed operators still satisfy the completeness condition $\sum_{m=0}^M {V_m}(\theta)^\dagger V_m(\theta)=I$ and thus define a valid CPTP map. The original dynamics is recovered when $\theta=0$. This perturbation is associated with a virtual acceleration of the dynamics and thereby enhances the environmental jumps [see the example with a specific $U_{SE}$ in Appendix~\ref{section:appendix:derivation-U-SE-example}], analogous to the conventional derivation of classical TURs via virtual perturbations \cite{Dechant2020-im,Hasegawa2019-fs,Koyuk2020-qm}. For any observable $\mathcal{F}$ on \textit{S}+\textit{E}, the quantum Cramér--Rao inequality states 
\begin{align}
    \frac{{\rm Var}[\mathcal{F}]}{(\partial_\theta \expval{\mathcal{F}}_{\theta=0})^2}\geq \frac{1}{J},\label{eq:end-matter-derivation-cramerrao}
\end{align}
where $J$ is the quantum Fisher information, defined below \cite{Paris2009-lx,Liu2019-iq}.

We now show that the terms $\partial_\theta \expval{\mathcal{F}}_{\theta=0}$ and $J$ in Eq.~\eqref{eq:end-matter-derivation-cramerrao} correspond to physical quantities under our definition of $\theta$. The derivative of the expectation value of the observable $\mathcal{F}$ with respect to $\theta$ is given by:
\begin{align}
    \partial_\theta \expval{\mathcal{F}}_\theta
    &= \qty(\sum_{k}\bra{\Psi_{S}(0)}\frac{dV_k(\theta)^\dagger}{d\theta}\otimes \bra{k_E})\mathcal{F}\nonumber\\
    &\quad\times\qty(\sum_{l} V_l(\theta)\ket{\Psi_{S}(0)}\otimes\ket{l_E})+{\rm h.c.}\label{eq:main-text-derivativeF}
\end{align}

For $k>0$, the derivative of $V_k(\theta)$ evaluated at $\theta=0$ is
\begin{align}
     \eval{\frac{dV_k(\theta)}{d\theta}}_{\theta=0}= \frac{1}{2}V_k.\label{eq:derivative-Vk}
\end{align}
For $k=0$, we use the spectral decomposition of $\sum_{m>0}^M V_m^\dagger V_m$:
\begin{align}
    \sum_{m>0}^M V_m^\dagger V_m = \sum_n \zeta_n \Pi_n,\label{eq:end-matter-spectral-decomposition-Vm}
\end{align}
where $\zeta_n$ is the eigenvalue of $\sum_{m>0}^M V_m^\dagger V_m$ and $\Pi_n$ is the projection operator onto the eigenspace corresponding to $\zeta_n$. Thus, we can write $V_0(\theta)$ as
\begin{align}
    V_0(\theta) &= U_0\sum_n \sqrt{1-e^{\theta}\zeta_n}\Pi_n.\label{eq:end-matter-spectral-decomposition-V0}
\end{align}
Using this equation, we can calculate the derivative with respect to $\theta$ as
\begin{align}
    \eval{\dv{V_0(\theta)}{\theta}}_{\theta=0}&= \frac{1}{2}U_0\sum_n \sqrt{1-\zeta_n}\Pi_n - \frac{1}{2}U_0\sum_n \frac{1}{\sqrt{1-\zeta_n}}\Pi_n\nonumber\\
    &= \frac{1}{2}V_0 - \frac{1}{2}V_0^{\dagger^{-1}}.\label{eq:derivative-V0}
\end{align}
By substituting these derivatives into Eq.~\eqref{eq:main-text-derivativeF}, we obtain $\partial_\theta \expval{\mathcal{F}}_{\theta=0}=\expval{\mathcal{F}}-\mathcal{C}_\mathcal{F}$, where $\mathcal{C}_\mathcal{F}$ is defined in Eq.~\eqref{eq:inherent-dynamics-contribution}. 

Next, the quantum Fisher information for the CPTP map is given by \cite{Escher2011-ja}
\begin{align}
    J(\theta) = 4[\ev**{H_1(\theta)}{\Psi_{S}(0)}-\ev**{H_2(\theta)}{\Psi_{S}(0)}^2],\label{eq:main-texti-QFI}
\end{align}
where
\begin{align}
H_1(\theta)&=  \sum_{m=0}^M \dv{V_m^\dagger(\theta)}{\theta}\dv{V_m}{\theta},\\
H_2(\theta)&= i\sum_{m=0}^M\dv{V_m^\dagger(\theta)}{\theta}V_m(\theta).
\end{align}

Using the derivatives $dV_k(\theta)/d\theta$ in Eqs.~\eqref{eq:derivative-Vk} and \eqref{eq:derivative-V0}, $H_1$ and $H_2$ are reduced to
\begin{align}
    H_1(\theta=0)&=\frac{1}{4}\qty(\qty({V_0}^\dagger V_0)^{-1}-I),\\
    H_2(\theta=0)&=0.
\end{align}
Therefore, we obtain
\begin{align}
    J(\theta=0)=\expval*{({V_0}^\dagger V_0)^{-1}}_{\rho_S(0)}-1=\mathcal{A}.
\end{align}
Combining these results, we arrive at the general QTUR [Eq.~\eqref{eq:GeneralTUR}].

\subsection{Equality condition of the general QTUR}
The equality condition of the general QTUR follows from that of the quantum Cramér--Rao inequality [Eq.~\eqref{eq:end-matter-derivation-cramerrao}], where the equality is achieved by the symmetric logarithmic derivative $\mathcal{L}$ with respect to $\theta$ \cite{Hotta2004-px}:
\begin{align}
    \mathcal{L}=\rho_{SE}(T)-\frac{1}{2}(\ketbra*{\Psi_{SE}(T)}{\Psi_{SE}(0)}{(V_0}^{-1}\otimes I_E)+\text{h.c.}),\label{eq:optimal-observable-L}
\end{align}
when $\rho_S(0)$ is pure [see also Eq.~\eqref{eq:appendix:logarithmic-derivative-L} in Appendix~\ref{section:appendix:derivation-of-generalQTUR}]. The eigenstates of $\mathcal{L}$ non-trivially depend on the initial and final states of $S$ and $E$, and the CPTP process. In particular, $\mathcal{L}$ generally requires collective measurements over $S+E$. In the first demonstration [Section~\ref{section:demonstration-I}], we aim to saturate the general QTUR by implementing the optimal observable $\mathcal{L}$ on qubits, utilizing the measurement capability of quantum processors.

\subsection{Application: quantum time correlator}\label{section:application-correlator}
The broad applicability of the general QTUR [Eq.~\eqref{eq:GeneralTUR}] enables us to establish trade-off relations beyond the precision of Hermitian observables. In this section, we consider the relationship between the thermodynamic quantity $\mathcal{A}$  and quantum time correlator, which is a topic of increasing interest in trade-off research \cite{Hasegawa2024-ki,Ohga2023-fk,Van_Vu2024-wh}. We later verify the derived bounds in the second demonstration [Section~\ref{section:demonstration-II}].

For a given CPTP map $\Phi$, the two-point time correlator of a quantum system is given by
\begin{align}
    R(T) = \langle \mathcal{H}(T)\mathcal{G}\rangle_{\rho_{S}(0)}\label{eq:quantum-time-correaltor}    
\end{align}
where $\mathcal{G}$ and $\mathcal{H}$ are \textit{unitary} and Hermitian observables, and $\mathcal{H}(T)=\sum_{m} V_m^{\dagger} \mathcal{H}V_m$ represents the Heisenberg picture of $\mathcal{H}$ after the process $\Phi$ \cite{Kubo1957-wb,Zwanzig1965-bf}. We investigate $R(T)$ through the general QTUR, which is an unexplored application that expands the scope of TURs.

$R(T)$ is measured using the Hadamard test with ancilla qubit $S'$ \cite{Somma2002-po}, as exemplified by the system in Fig.~\ref{fig:illustration}b [see Appendix~\ref{section:appendix:details-circuits-for-Fig1b} for details]. Due to its general applicability, the general QTUR remains valid throughout this entire measurement process. The correspondence between the elements involved in evaluating $R(T)$ and those in the generic formulation (depicted in Fig.~\ref{fig:illustration}a) is color-coded: the effective initial state $\rho_{S'S}^\mathcal{G}$ (light blue region) and the effective observable $\mathcal{F}_{\rm corr}$ satisfying $\expval*{\mathcal{F}_{\rm corr}}_{\rho_S(T)}= {\rm Re}[R(T)]$ (purple region). Although $R(T)$ is not itself the expectation of a physical (i.e., Hermitian) observable, its measurement fluctuations still obey the QTUR, with the associated survival activity $\mathcal{A}$:
\begin{align}
\frac{{\rm Var}[{\rm Re}[R(T)]]}{({\rm Re}[R(T)] - \mathcal{C}_{\mathcal{F}_{\rm corr}})^2} \geq \frac{1}{\mathcal{A}}.\label{eq:generalQTUR-for-time-correlator}
\end{align}

Furthermore, the general QTUR not only constrains the precision but also directly leads to a trade-off relation for the value of $R(T)$ itself:
\begin{equation}
    |{\rm Re}[R(T)] - \mathcal{C}_{\mathcal{G},\mathcal{H}}| \leq \sqrt{\mathcal{A}}, \label{eq:timeCorrelationBound}
\end{equation}
where the baseline $\mathcal{C}_{\mathcal{G},\mathcal{H}}$ is given by $\mathcal{C}_{\mathcal{G},\mathcal{H}} = {\rm Re}[\expval*{{V_0}^{-1}\mathcal{H}V_0\mathcal{G}}_{\rho_S(0)} + \expval*{\mathcal{G}{V_0}^{-1}\mathcal{H}V_0}_{\rho_S(0)}]/2$. Equation~\eqref{eq:timeCorrelationBound} shows that the survival activity bounds time correlation generated by environmental interaction, represented as the deviation of $R(T)$ from the correlation dictated by the system's intrinsic dynamics: we show that, in the weak coupling regime where ${V_0}^\dagger V_0 = I_S - \epsilon$, $\mathcal{C}_{\mathcal{G},\mathcal{H}} = {\rm Re}[\expval*{{V_0}^\dagger \mathcal{H}V_0 \mathcal{G}}_{\rho_S(0)}] + \mathcal{O}(\epsilon)$. The derivation of Eqs.~\eqref{eq:generalQTUR-for-time-correlator} and \eqref{eq:timeCorrelationBound} are given in Appendices \ref{section:appendix:derivation-of-RT-precision-bound} and \ref{section:appendix:derivation-of-RT-tradeoff}, respectively.

\section{Verification methodology}\label{section:verification-methodology}
In this section, we introduce a quantum-computer-based verification methodology, which will later be applied to two demonstrations: (i) testing the general QTUR [Eq.~\eqref{eq:GeneralTUR}] with the optimal observable, and (ii) verifying trade-offs in quantum time correlators [Eqs.~\eqref{eq:generalQTUR-for-time-correlator} and \eqref{eq:timeCorrelationBound}]. We first discuss the features of quantum computers as a demonstration platform. We then present a protocol to measure the key quantities in the bounds, namely the survival activity $\mathcal{A}$ and the inherent-dynamics contribution $\mathcal{C}_\mathcal{F}$. Finally, to implement the protocol on current quantum processors, we develop a circuit optimization technique tailored to the general QTUR.

\subsection{Quantum computer as a thermodynamic testbed}\label{section:main:quantum-computer}
Our demonstrations are performed on IBM’s superconducting quantum processor ``Heron'' (ibm\_torino)~\cite{Ibm_undated-ip}. The device characteristics, such as coherence times and error rates, are summarized in Appendix~\ref{section:appendix:IBM}. In our setup, transmon qubits~\cite{Koch2007-qf} constitute the system and environment.  We examine their fluctuations following unitary evolution $U_{SE}$, implemented using quantum gates. The key coupling dynamics in our demonstrations is realized by a controlled-RY gate with rotation angle $\pi\gamma$ [see Figs.~\ref{fig:Fopt}a and \ref{fig:illustration}b for details]. As shown in Appendix~\ref{section:appendix:derivation-U-SE-example}, the survival activity $\mathcal{A}$ is theoretically proportional to $\gamma^2$ in the weak-coupling regime $\gamma \ll 1$; accordingly, we refer to $\gamma$ as the \textit{interaction strength} hereafter.

Real qubits and gates are far from the idealized versions often implicitly assumed in quantum algorithms. In practice, they exhibit ``imperfections'' such as infidelity and dissipation stemming from thermodynamic constraints. These limitations originate in the physical implementation of quantum operations. For example, the controlled-RY gate on the processor is constructed from controlled-Z gates mediated by a tunable coupler (requiring additional single-qubit rotations) \cite{McKay2023-tj}. Thus, thermodynamic laws, notably the general QTUR, are expected to manifest in execution outcomes. This principle forms the core of our verification methodology.

\subsection{Measurement protocol for $\mathcal{A}$ and $\mathcal{C}_\mathcal{F}$}\label{section:main:measure-A-and-CF}
The challenge in empirically verifying the general QTUR [Eq.~\eqref{eq:GeneralTUR}] lies in the measurement of $\mathcal{A}$ and $\mathcal{C}_\mathcal{F}$, as their definitions involve ${V_0}^{-1}$, an operator whose inversion is generally intractable. To overcome this, we first introduce a generic experimental protocol for measuring these quantities, and then present a circuit simplification technique suitable for current quantum processors in the next subsection.

As a key step, we avoid matrix inversion by employing a Neumann series expansion of $(V_0^\dagger V_0)^{-1}$, which rewrites the inverse in terms of experimentally accessible quantities. Using the completeness condition of the Kraus operators $\{V_m\}$, we define an operator $\epsilon=\sum_{m>0}V_m^\dagger V_m=I-V_0^\dagger V_0$. The Neumann series expansion of $(V_0^\dagger V_0)^{-1}$ is given by
\begin{align}
    (V_0^\dagger V_0)^{-1} = (I-\epsilon)^{-1} = \sum_{n=0}^\infty \epsilon^n = \sum_{n=0}^\infty (I-V_0^\dagger V_0)^n. \label{eq:appendix:V0-expansion}
\end{align}
Applying the binomial expansion, we obtain
\begin{align}
    (V_0^\dagger V_0)^{-1} = \sum_{n=1}^\infty \sum_{k=0}^n (-1)^k\binom{n}{k}(V_0^\dagger V_0)^{k}+1.
\end{align}
Using this expansion, the survival activity $\mathcal{A}$ can be expressed as
\begin{align}
    \mathcal{A} &= \lim_{N\rightarrow\infty}\sum_{n=0}^N (-1)^n\binom{N+1}{n+1}\expval*{(V_0^{\dagger} V_0)^{n}}_{\rho_{S}(0)}-1,\label{eq:appendix:xi-expansion}
\end{align}
where each expectation value $\expval*{(V_0^{\dagger} V_0)^{n}}_{\rho_{S}(0)}$ can be measured experimentally. For weak system-environment interactions, as in our demonstrations, the series can be truncated. In the following, we set $N = 1$ and approximate $\mathcal{A} \approx 1 - p_0$, where $p_0=\expval*{{V_0}^{\dagger} V_0}_{\rho_{S}(0)}$ denotes the probability that the environment remains in its initial state. Therefore, $\mathcal{A}$ can be empirically determined by measuring the environmental qubits after the application of $U_{SE}$. Similarly, $\mathcal{C}_\mathcal{F}$ is approximated as $\mathcal{C}_\mathcal{F} \approx 2C_1 - C_2$, where
\begin{align}
    C_1&={\rm Re}[\expval*{(V_0^{\dagger}\otimes I_E)\mathcal{F}U_{SE}}_{\rho_{SE}(0)}],\label{eq:main-text-C1}\\
    C_2&={\rm Re}[\expval*{({V_0}^{\dagger} V_0{V_0}^{\dagger}\otimes I_E)\mathcal{F}U_{SE}}_{\rho_{SE}(0)}].\label{eq:main-text-C2}
\end{align}

In Appendix~\ref{section:appendix:measurement-of-A-theory}, we provide a circuit implementation for the higher-order estimation of $\mathcal{A}$ using $U_{SE}$ and $U_{SE}^\dagger$. This protocol is generic, as it does not rely on microscopic details of the system or its time evolution, and all components are implementable on other platforms. Therefore, identifying $\mathcal{A}$ using this protocol on a quantum processor would imply that $\mathcal{A}$ is physically measurable.

\subsection{Circuit optimization}\label{section:demonstration-I-circuit-optimzation}
\begin{figure}[t]
    \centering
    \subfloat[]{\includegraphics[width=0.6\columnwidth]{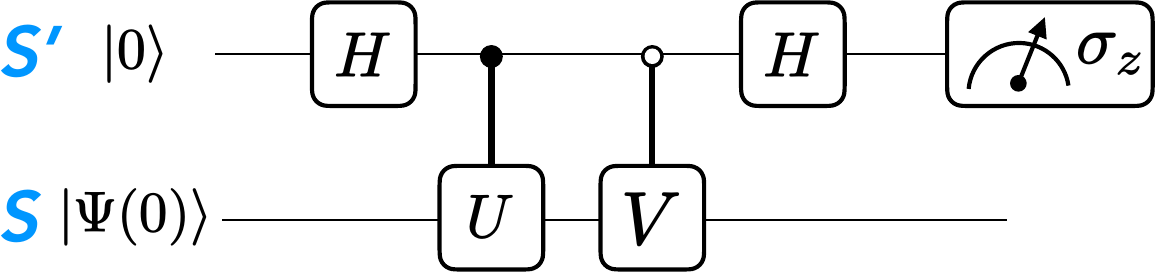}}\\
    \subfloat[]{\includegraphics[width=0.7\columnwidth]{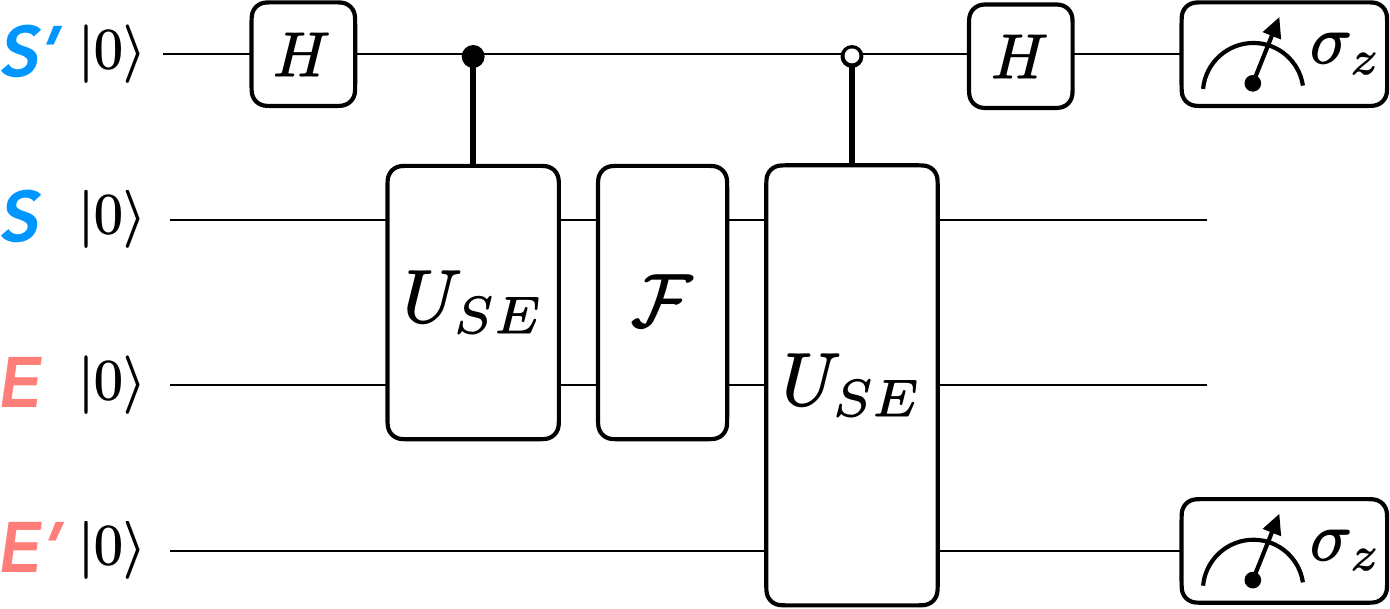}}\\
    \subfloat[]{\includegraphics[width=0.6\columnwidth]{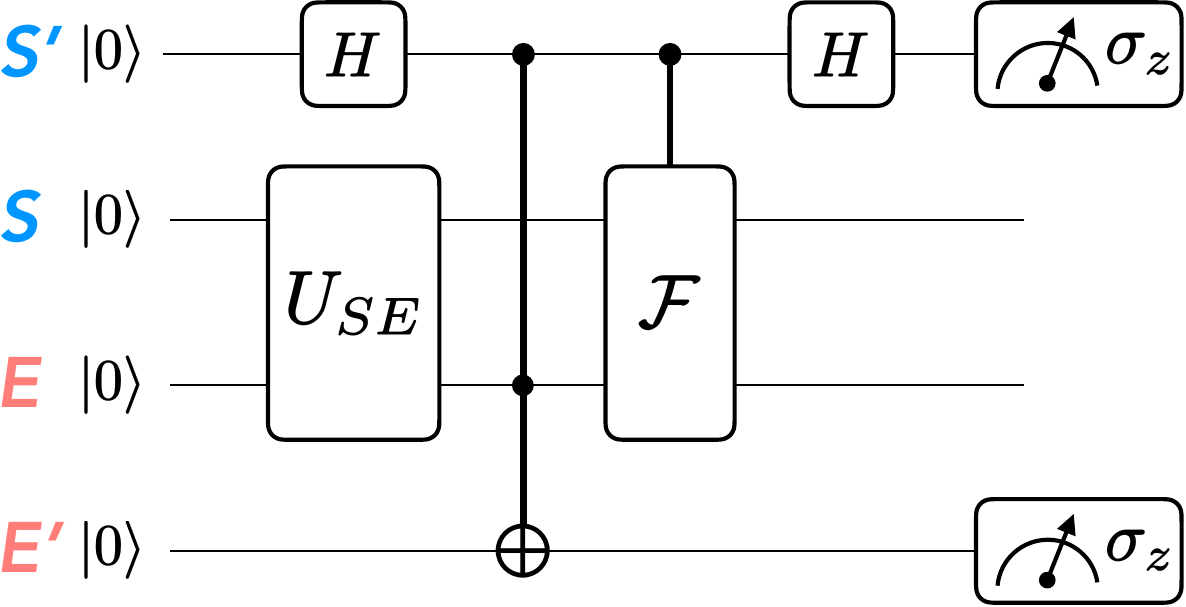}}
    \caption[]{\label{fig:main:circuit-for-C1}Circuits for measuring the inherent-dynamics contribution $\mathcal{C}_\mathcal{F}$. (a) Hadamard test for evaluating a generic correlator ${\rm Re}[\bra{\Psi(0)}V^\dagger U\ket{\Psi(0)}]$. The white control indicates that the target unitary is applied when the control qubit is in the $\ket{0}$ state. (b) Straightforward implementation of the Hadamard test circuit for measuring $C_1$. (c) Optimized circuit for $C_1$, where the controlled-CNOT (Toffoli) gate is implemented approximately.
    }
\end{figure}
In principle, the above protocol can be represented as quantum circuits, though not executable on current hardware. Specifically, the correlators $C_1$ and $C_2$ in Eqs.~\eqref{eq:main-text-C1} and \eqref{eq:main-text-C2} are measured using the Hadamard tests~\cite{Somma2002-po} [see Fig.~\ref{fig:main:circuit-for-C1}a and Appendix~\ref{section:appendix:C-F} for details]. In particular, $C_1$ is obtained by  sequentially applying $U_{SE}$, $\mathcal{F}$, and $V_0$ controlled by an ancillary qubit $S'$, as shown in Fig.~\ref{fig:main:circuit-for-C1}b. Here, the dynamics $V_0$ is implemented by applying $U_{SE}$ and postselecting the environmental qubit $E'$ in the state $\ket{0}$, where $E'$ is a replica of the original environmental qubit $E$. For $C_2$, the circuit involves the controlled versions of $U_{SE}$, $\mathcal{F}$, and $V_0 {V_0}^\dagger V_0$, resulting in a longer sequence than that for $C_1$.

However, implementing multi-qubit controlled gates requires deep circuits due to hardware constraints such as restricted native gate sets and limited qubit connectivity \cite{Kusyk2021-rk}. For instance, decomposing logical gates into native physical gates necessitates swap operations when acting on physically distant qubits, with each swap comprising three CNOT gates. As a result, current devices, which are highly susceptible to gate errors and decoherence, cannot execute such deep circuits reliably.

To address this issue, we develop a circuit optimization that leverages the algebraic structure of $C_1$ and $C_2$ to reduce the number of costly controlled operations. Specifically, due to the symmetry between $U_{SE}$ and $V_0$ in the definition of $C_1$ [Eq.~\eqref{eq:main-text-C1}], the corresponding controlled operations, namely controlled-$U_{SE}$ and controlled-$V_0$, can be unified into a single uncontrolled operation as follows. We rewrite Eq.~\eqref{eq:main-text-C1} as
\begin{align}
    C_1&={\rm Re}[\expval*{U_{SE}^\dagger(I_S\otimes \ketbra{0}{0})\mathcal{F}U_{SE}}_{\rho_{SE}(0)}].\label{eq:main-text-C1-rewrite}
\end{align}
In the Hadamard test, Hermitian-conjugate symmetric operations, for example $U_{SE}$ in Eq.~\eqref{eq:main-text-C1-rewrite}, do not require control by the ancilla. Therefore, the controlled-$U_{SE}$ can be avoided by instead introducing the projection $I_{S}\otimes\ketbra{0}{0}$, which can be implemented using a controlled-CNOT (Toffoli) gate, depicted in Fig.~\ref{fig:main:circuit-for-C1}c. 

We further reduce the depth of the improved circuit in Fig.~\ref{fig:main:circuit-for-C1}c by employing approximate Toffoli gates~\cite{He2017-dw} developed in quantum computing. These gates simplify implementation by relaxing the exactness of certain input-output mappings. Remarkably, they \textit{behave identically to full Toffoli gates in our setup}, where the environmental qubit is initialized in the $\ket{0}$ state, as shown in Fig.~\ref{fig:main:circuit-for-C1}c. 

Thanks to these optimization techniques, the circuit depth in each demonstration is reduced by roughly an order of magnitude, with details provided therein. A similar optimization is applicable to $C_2$, which is discussed in Appendix~\ref{section:appendix:circuits-for-C-F-opt}. Furthermore, in the second demonstration on the quantum time correlator [Section~\ref{section:demonstration-II}], we achieve additional optimization by exploiting the structures of $C_1$ and $C_2$ that are specific to the target observables.

\section{Demonstrations}\label{section:demonstrations}
As described in the Introduction, a fundamental QTUR has to be empirically accessible to offer predictions on experiments. To this end, we conduct two demonstrations using quantum computers. First, in Section~\ref{section:demonstration-I}, we investigate the measurability of the cost function and empirical tightness of the general QTUR [Eq.~\eqref{eq:GeneralTUR}] by implementing the optimal observable. Then, we explore trade-offs on quantum time correlators [Eqs.~\eqref{eq:generalQTUR-for-time-correlator} and \eqref{eq:timeCorrelationBound}] in Section~\ref{section:demonstration-II} to show the applicability of our theoretical framework and verification methodology. Through these demonstrations, we highlight the empirical relevance of the general QTUR and the impact of the quantum-computer-based verification.

\subsection{Optimal observable}\label{section:demonstration-I}
\begin{figure}[t]
    \centering
    \subfloat[]{\includegraphics[width=0.65\columnwidth]{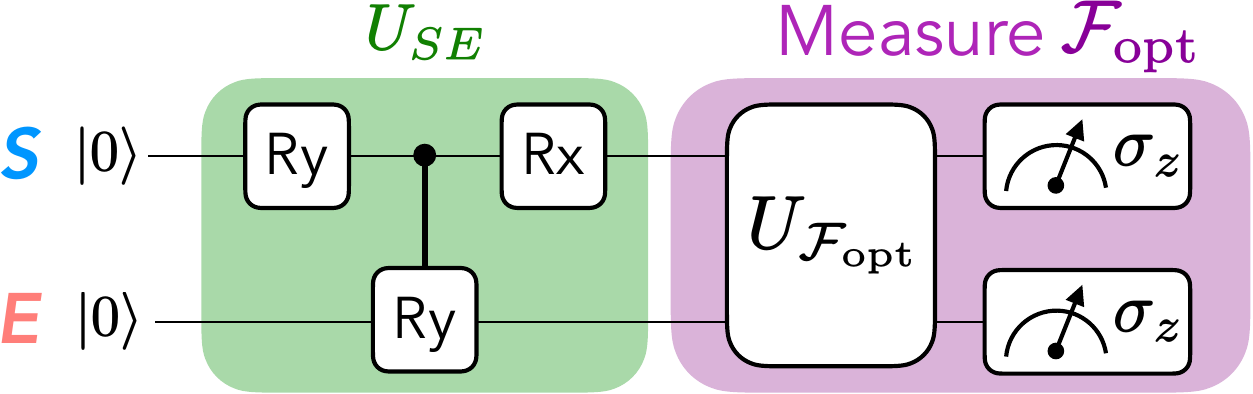}}\\
    \subfloat[]{\includegraphics[width=0.7\columnwidth]{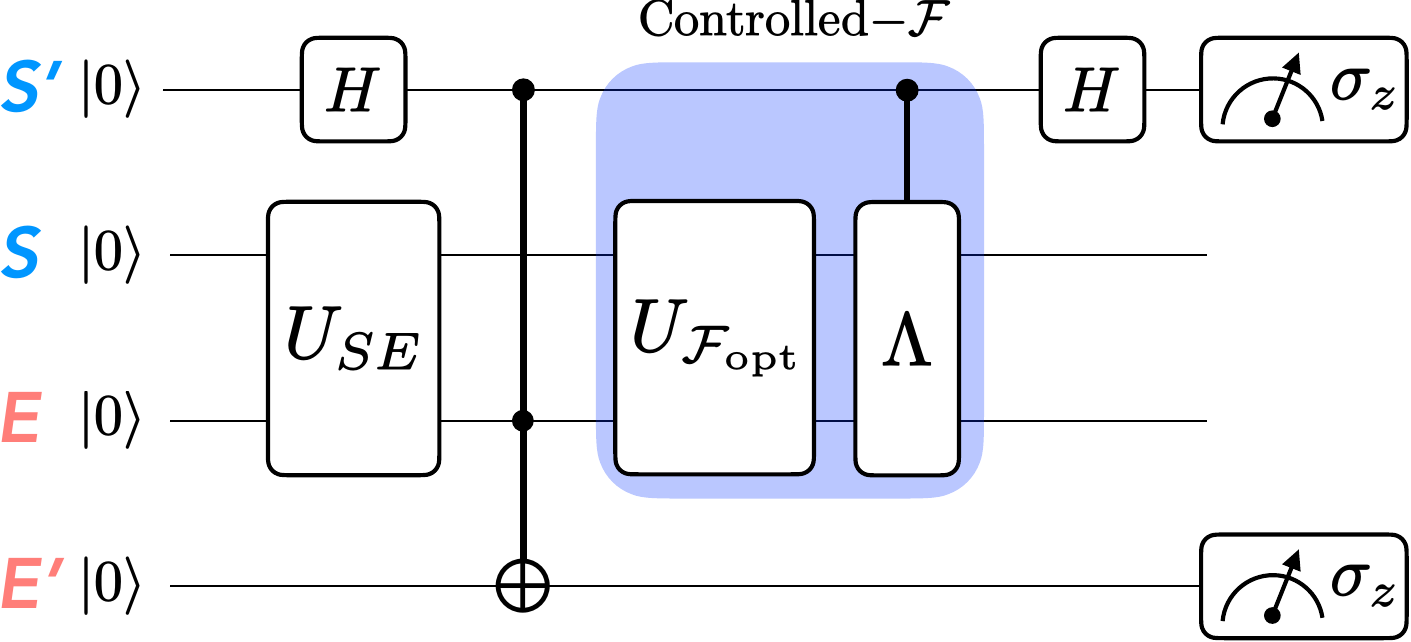}}
    \caption[]{\label{fig:Fopt}Circuits for the demonstration of the optimal observable $\mathcal{F}_{\rm opt}$. (a) Circuit implementing the unitary $U_{SE}$ and the optimal observable $\mathcal{F}_{\rm opt}$. The rotation angles of the gates are configuration-dependent. In particular, the controlled-RY gate applies a rotation by angle $\pi\gamma$, where $\gamma$ denotes the interaction strength. (b) Optimized circuits for measuring $C_1$, where the controlled-CNOT (Toffoli) gate is implemented approximately. The controlled-$\mathcal{F}$ gate is decomposed into $U_{\mathcal{F}_{\rm opt}}$ and a controlled-$\Lambda$ gate.
    }
\end{figure}
\subsubsection{Aim and design}\label{section:demonstration-I-goals-and-design}
In this demonstration, we aim to establish the practical significance of the general QTUR [Eq.~\eqref{eq:GeneralTUR}]. Specifically, we examine:
\begin{enumerate}
    \item Empirical measurability of the thermodynamic cost, i.e., survival activity $\mathcal{A}$, and the correction term $\mathcal{C}_\mathcal{F}$.
    \item Validity of the bound in real systems beyond theoretical frameworks.
    \item Empirical tightness on current quantum devices, particularly the attainability of the bound.
\end{enumerate}
We investigate these aspects on the quantum processor by assessing the fluctuations of the optimal observable that theoretically saturates the bound. While no previous study has realized the saturation of a general QTUR, we achieve this by leveraging the unique measurement flexibility of quantum computers.

Theoretically, the optimal observable $\mathcal{F}_{\rm opt} = \mathcal{L}$ [Eq.~\eqref{eq:optimal-observable-L}] minimizes thermodynamic inefficiency, defined as the following product of the variation and the thermodynamic cost:
\begin{align}
    \frac{{\rm Var}[\mathcal{F}]}{(\langle \mathcal{F} \rangle -\mathcal{C}_\mathcal{F})^2}\cdot\mathcal{A},
\end{align}
which becomes unity when the bound is saturated. In other words, $\mathcal{F}_{\rm opt}$ maximizes the precision for a given thermodynamic cost $\mathcal{A}$. Empirical saturation is significant for the study of QTURs, as any residual gap would imply that the precision could be further improved without additional thermodynamic cost, contradicting the premise of a true fundamental limit. Thus, observing saturation of the general QTUR would provide compelling evidence for the universal principle that enhancing precision necessarily entails a thermodynamic cost.

Measuring $\mathcal{F}_{\rm opt}$ is difficult on conventional platforms because it requires entangled measurements over the composite system $S+E$. In contrast, quantum computers can perform such measurements. We consider a two-qubit system for $S+E$, initialized in $\ket{0} \otimes \ket{0}$ and evolved under $U_{SE}$ which is parameterized by the interaction strength $\gamma$ as illustrated in Fig.~\ref{fig:Fopt}a. Given $U_{SE}$, we numerically compute $\mathcal{F}_{\rm opt}$ and a unitary operator $U_{\mathcal{F}_{\rm opt}}$ that diagonalizes it as
\begin{align}
    \mathcal{F}_{\rm opt}=U_{\mathcal{F}_{\rm opt}}^\dagger \Lambda U_{\mathcal{F}_{\rm opt}},\label{eq:Fopt-diagonalization}
\end{align}
where $\Lambda$ is a diagonal matrix. By implementing $U_{\mathcal{F}_{\rm opt}}$ on $S+E$ as a quantum circuit, we can access $\mathcal{F}_{\rm opt}$ through the measurement of the diagonal operator $\Lambda$. Notably, any two-qubit unitary can be implemented using only three CNOT gates and several single-qubit gates~\cite{Vatan2004-zk}, rendering this approach feasible on current processors. See Appendix~\ref{section:appendix:Fopt-system} for details of the parameterization of $U_{SE}$ and  the measurement of $\mathcal{F}_{\rm opt}$.

The involved quantities $\mathcal{A}$ and $\mathcal{C}_\mathcal{F}$ are measured using the protocol introduced in Section~\ref{section:main:measure-A-and-CF}. We then apply the circuit optimization technique presented in Section~\ref{section:demonstration-I-circuit-optimzation} with the target $U_{SE}$ and $\mathcal{F}_{\rm opt}$, yielding the optimized circuit shown in Fig.~\ref{fig:Fopt}b (for $C_1$). In the latter part of the circuit, the implementation of controlled-$\mathcal{F}_{\rm opt}$ in the Hadamard test requires only the controlled-$\Lambda$, owing to Eq.~\eqref{eq:Fopt-diagonalization}. As a result of the optimization, the circuit depth for $C_1$ is reduced by an order of magnitude compared to the straightforward implementation: the median depth decreases from 539 to 48.5. A similar optimization is applicable to $C_2$, leading to a depth reduction from 720 to 90. See Appendices~\ref{section:appendix:circuits-for-C-F-opt} and \ref{section:appendix:depth-evaluation-F-opt} for details of circuit implementations and depth evaluation, respectively. 

\subsubsection{Empirical results}
\begin{figure*}[t]
    \centering
    \includegraphics[width=0.95\textwidth]{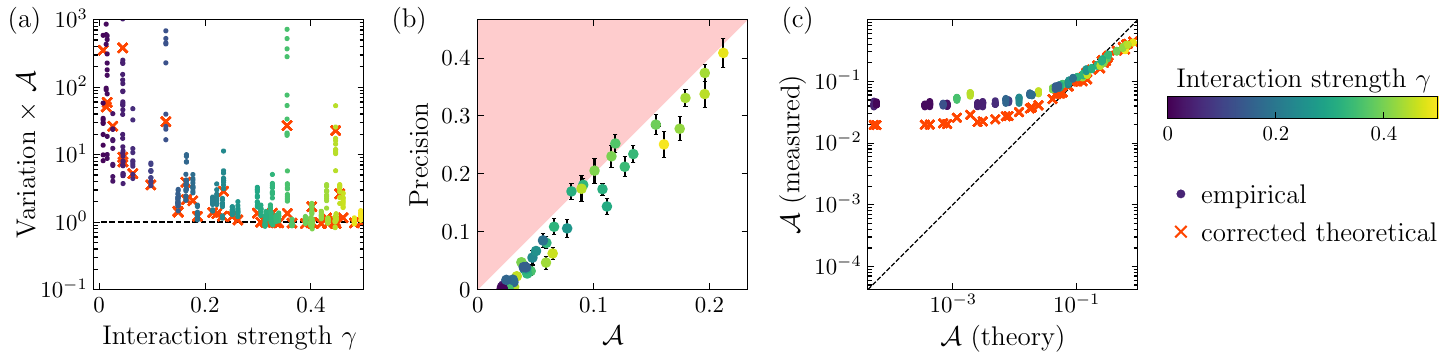}
    \caption[]{\label{fig:experiment-result-Fopt}Demonstration of the general QTUR using the optimal observable $\mathcal{F}_{\rm opt}$ on IBM's quantum processor (ibm\_torino). Data points correspond to different system configurations, colored by the interaction strength $\gamma$. (a) Verification of the general QTUR's equality condition. Empirical and thermally-corrected theoretical values (with the Neumann series approximation) of the thermodynamic inefficiency (the variation ${\rm Var}[\mathcal{F}_{\rm opt}]/(\expval*{\mathcal{F}_{\rm opt}}-\mathcal{C}_{\mathcal{F}_{\rm opt}})^2$ multiplied by the survival activity $\mathcal{A}$) versus $\gamma$. (b) Cost-Precision trade-off. Precision (the inverse of the variation, i.e., $(\expval*{\mathcal{F}_{\rm opt}}-\mathcal{C}_{\mathcal{F}_{\rm opt}})^2/{\rm Var}[\mathcal{F}_{\rm opt}]$) versus $\mathcal{A}$. The shaded region indicates the domain forbidden by the general QTUR. Error bars represent one standard deviation. (c) Comparison of empirical and thermally-corrected theoretical values of $\mathcal{A}$ with their theoretical counterparts. The dashed line represents $y=x$.  Settings: 50 circuits with different parameters were constructed, each undergoing 8,000 measurements, repeated 10 times.
    }
\end{figure*}
We perform demonstrations with varying $U_{SE}$, particularly focusing on the interaction strength $\gamma$. For each configuration, we collect 8,000 measurement shots to evaluate expectation values and repeat this procedure 10 times. The resulting data are shown in Fig.~\ref{fig:experiment-result-Fopt}. In Fig.~\ref{fig:experiment-result-Fopt}a, the empirical thermodynamic inefficiency, ${\rm Var}[\mathcal{F}_{\rm opt}]/(\expval{\mathcal{F}_{\rm opt}}-\mathcal{C}_{\mathcal{F}_{\rm opt}})^2 \times \mathcal{A}$, exceeds unity, consistent with the general QTUR [Eq.~\eqref{eq:GeneralTUR}]. For stronger interactions ($\gamma>0.3$), the inefficiency nearly saturates to 1, marking the first empirical observation of QTUR saturation. To illustrate this as a cost-precision trade-off, we plot the precision---defined as the reciprocal of the variation, i.e., $(\expval{\mathcal{F}_{\rm opt}}-\mathcal{C}_{\mathcal{F}_{\rm opt}})^2/{\rm Var}[\mathcal{F}_{\rm opt}]$---against $\mathcal{A}$ in Fig.~\ref{fig:experiment-result-Fopt}b. The data demonstrate the thermodynamic precision limit, which underscores the principle that precision cannot be improved without incurring greater cost. The observed saturation also indicates the sharpness of the bound, suggesting that current quantum devices are sufficiently accurate to capture the precision limit. 

Fig.~\ref{fig:experiment-result-Fopt}c displays the measured survival activity $\mathcal{A}$. For large $\mathcal{A}$, the empirical values (colored points) closely follow the theoretical prediction (dashed line). Since our protocol is generic and applicable to other platforms, this result confirms that $\mathcal{A}$ is empirically accessible. Altogether, these findings in Figs.~\ref{fig:experiment-result-Fopt}a--\ref{fig:experiment-result-Fopt}c suggest that measuring $\mathcal{A}$ alone allows one to estimate the  system's achievable precision without testing various observables. In Appendix~\ref{section:appendix:Fopt-empirical-results}, we validate that the influence of truncating the series in Eq.~\eqref{eq:appendix:xi-expansion} is insignificant compared to the fluctuations of the empirical data.

In contrast, for small $\gamma$ (i.e., small $\mathcal{A}$), the empirical results deviate from theoretical predictions: in Fig.~\ref{fig:experiment-result-Fopt}a, the inefficiency differs from unity, and in Fig.~\ref{fig:experiment-result-Fopt}c, the measured values of $\mathcal{A}$ exceed the theoretical estimates by several orders of magnitude. We attribute this discrepancy primarily to the finite-temperature effect of the qubits in the processor. While our theoretical framework assumes that $E$ is initially in a pure state (though it can be generalized to thermal environments, as discussed in Appendix~\ref{section:appendix:generalization-to-thermal-environments}), the actual device operates at a finite effective temperature. Calibration data show that the qubits exhibit approximately 2\% excited-state populations at equilibrium, including contributions from readout errors. When this thermal effect is incorporated into the theoretical model, the corrected predictions (orange crosses in Figs.~\ref{fig:experiment-result-Fopt}a and \ref{fig:experiment-result-Fopt}c) show excellent agreement with the empirical results. Detailed numerical analysis is provided in Appendix~\ref{section:appendix:analyze-errors-Fopt}.

These results indicate that finite temperature can significantly influence the precision bound determined by the survival activity $\mathcal{A}$, yet the general QTUR retains quantitative validity in certain regimes even under such thermal conditions. Assuming a typical qubit frequency of 5~GHz, we estimate an effective temperature in the tens of millikelvin range, which is consistent with Refs.~\cite{Solfanelli2021-rc,Buffoni2022-jf,Bassman-Oftelie2024-nf}. Our analysis thus demonstrates both the robustness of the theory under realistic cryogenic conditions and the identification of parameter regimes (small $\gamma$) where the bound dictated by $\mathcal{A}$ may become less stringent.

\begin{figure}[t]
    \centering
    \includegraphics[width=\columnwidth]{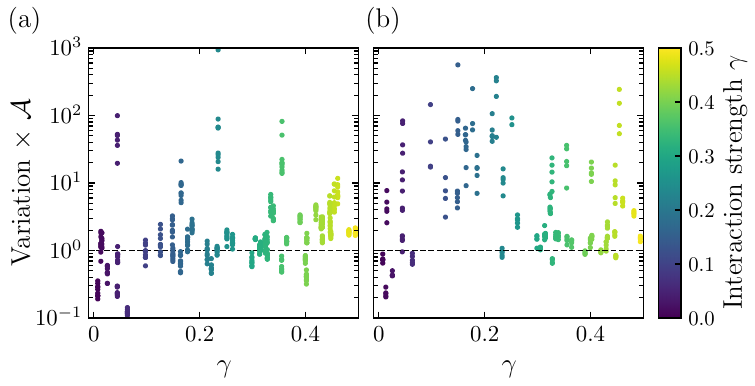}
    \caption[]{\label{fig:experiment-result-compare_nonoptimized_and_eagle}Reference empirical thermodynamic inefficiency, defined as the product of the variation ${\rm Var}[\mathcal{F}_{\rm opt}]/(\expval*{\mathcal{F}_{\rm opt}}-\mathcal{C}_{\mathcal{F}_{\rm opt}})^2$ and the survival activity $\mathcal{A}$. (a) Result from a straightforward (i.e., non-optimized) implementation of the circuits for $\mathcal{C}_{\mathcal{F}_{\rm opt}}$ on ibm\_torino. (b) Result from the optimized circuits on ibm\_sherbrooke, an Eagle-series processor that preceded the Heron. Settings: 50 circuits with different parameters were constructed, each undergoing 8,000 measurements, repeated 10 times for (a) and 5 times for (b).}
\end{figure}

Furthermore, to highlight the impact of circuit depth reduction, Fig.~\ref{fig:experiment-result-compare_nonoptimized_and_eagle}a shows that non-optimized circuits lead to a violation of the bound. This result indicates that our successful demonstration not only confirms the validity of the general QTUR, but also shows that our methodology makes current quantum hardware capable of verifying trade-off relations. Additionally, in Fig.~\ref{fig:experiment-result-compare_nonoptimized_and_eagle}b, we present results obtained using the previous-generation ``Eagle'' processor \cite{Ibm_undated-ip,McKay2023-tj,Kim2023-px}. The observed violation in this case indicates that improved hardware performance, such as that provided by Heron, is essential for a successful demonstration. In other words, these results underscore the need for high accuracy in verifying thermodynamic inequalities. Details of the empirical results for the straightforward implementation and for the Eagle processor are presented in Appendices~\ref{section:appendix:results-for-straightforward} and \ref{section:appendix:results-for-Eagle}, respectively.

\subsection{Quantum time correlator}\label{section:demonstration-II}
\subsubsection{Aim and design}
Our second demonstration showcases the applicability of our method and the general QTUR [Eq.~\eqref{eq:GeneralTUR}] by exploring thermodynamic trade-offs involving the quantum time correlator $R[T]$, as formalized in Eqs.~\eqref{eq:generalQTUR-for-time-correlator} and \eqref{eq:timeCorrelationBound}. This demonstration serves two purposes, motivated by the current lack of empirical studies on trade-offs for quantum time correlators. First, to validate the bounds in realistic systems. As detailed below, the system $S$ is larger and the dynamics $U_{SE}$ involve more gates than in the previous demonstration. Since the trade-off verification is sensitive to operational errors, it remains unclear whether such deeper circuits still obey the bounds. Second, to assess empirical tightness. Since $R[T]$ is not the optimal observable that saturates the QTUR, evaluating whether the survival activity $\mathcal{A}$ imposes a sharp constraint on $R[T]$ provides insight into the practical relevance of the bounds. 

To verify the bounds, we consider the composite system $S+E$ depicted in Fig.~\ref{fig:illustration}b, where the system $S$ comprises primary qubits $S_1$ and $S_2$, along with an ancillary qubit $S'$. Various initial states $\rho_S(0)$ are prepared using parameterized gates on $S_1$ and $S_2$. The dynamics $U_{SE}$ is implemented via parameterized entangling operations within $S$ and between $ S$ and $E$. In particular, the coupling is realized by a controlled-RY gate acting on $S_1$ and $E$, with rotation angle $\pi\gamma$, as in the previous demonstration [Fig.~\ref{fig:Fopt}]. The ancilla $S'$ is used to perform the Hadamard test, sandwiching $U_{SE}$, in order to measure the observable $\mathcal{F}_{\rm corr}$, which satisfies $\expval{\mathcal{F}_{\rm corr}}_{\rho_S(T)}={\rm Re}[R[T]]$. Equations~\eqref{eq:generalQTUR-for-time-correlator} and \eqref{eq:timeCorrelationBound} are evaluated using the same measurement data: specifically, the inherent-dynamics contribution $\mathcal{C}_{\mathcal{G},\mathcal{H}}$ is obtained via the relation $\mathcal{C}_{\mathcal{G},\mathcal{H}}=\mathcal{C}_{\mathcal{F}_{\rm corr}}$.

Most of the verification procedure follows that of the previous demonstration. The survival activity is measured with the protocol in Section~\ref{section:main:measure-A-and-CF} and circuit optimization is applied accordingly. However, we employ further circuit optimization for measuring $C_1$ and $C_2$, as described below.

\begin{figure*}[t]
    \centering
    \includegraphics[width=0.7\textwidth]{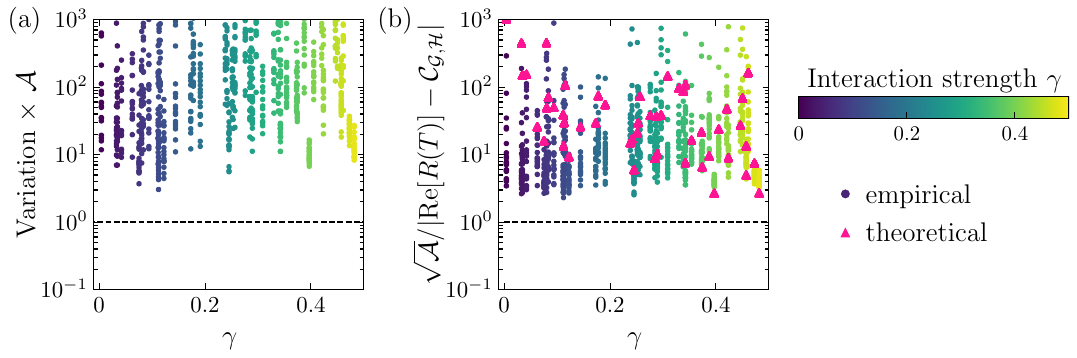}
    \caption[]{\label{fig:experiment-result-correlator}Demonstration of the trade-off relations involving the quantum time correlator $R(T)$ on ibm\_torino. Data points correspond to different system configurations, colored by the interaction strength $\gamma$. (a) Verification of the general QTUR for ${\rm Re}[R(T)]$. The thermodynamic inefficiency ${\rm Var}[{\rm Re[R(T)]}]/({\rm Re}[R(T)]-\mathcal{C}_{\mathcal{F}_{\rm corr}})^2\times \mathcal{A}$ versus $\gamma$. (b) Verification of the trade-off for the evolution of $R(T)$. Empirical and theoretical inefficiency given by the ratio between $\sqrt{\mathcal{A}}$ and $|{\rm Re}[R(T)]-\mathcal{C}_{\mathcal{G},\mathcal{H}}|$ versus $\gamma$. The data used are the same as in (a). Settings: 50 circuits with different parameters were constructed, each undergoing 4,000 measurements, repeated 20 times.}
\end{figure*}

\subsubsection{Additional circuit optimization}\label{section:demonstration-II-circuit-optimization}
A difficulty in measuring $C_1$ and $C_2$ is the use of controlled operations. In this demonstration, we further simplify the circuits for $C_1$ and $C_2$ by exploiting a theoretical observation. For a local observable $\mathcal{F}_S$ defined only on $S$ (excluding $E$), the quantity $\mathcal{C}_{\mathcal{F}_S}$ can be simplified as 
\begin{align}
        \mathcal{C}_{{\mathcal{F}_S}}= {\rm Re}[\expval{{V_0}^{-1}\mathcal{F}_S V_0}_{\rho_S(0)}],\label{eq:main-CFs}
\end{align}
involving only $V_0$ from the set $\{V_m\}$ composing $U_{SE}$ [see Appendix~\ref{section:appendix:simplification-of-CF-for-Fs} for the derivation]. This applies to the observable $\mathcal{F}_{\rm corr}$ considered in this demonstration. Consequently, the expression for $C_1$ and $C_2$ becomes
\begin{align}
    C_1&={\rm Re}[\expval*{V_0^{\dagger}\mathcal{F}_{\rm corr}V_0}_{\rho_{S}(0)}],\label{eq:main-text-C1-Fcorr}\\
    C_2&={\rm Re}[\expval*{{V_0}^{\dagger} V_0{V_0}^{\dagger}\mathcal{F}_{\rm corr}V_0}_{\rho_{S}(0)}].\label{eq:main-text-C2-Fcorr}
\end{align}

As described in Section~\ref{section:demonstration-I-circuit-optimzation}, the Hermitian-conjugate symmetry in Eqs.~\eqref{eq:main-text-C1-Fcorr} and \eqref{eq:main-text-C2-Fcorr} allows us to avoid controlled operations in the Hadamard test. Here, we note that the counterpart of $V_0V_0^\dagger$ in Eq.~\eqref{eq:main-text-C2-Fcorr} is $I=U_{SE}^\dagger U_{SE}$. As a result, even the Toffoli gate in the previously optimized circuit in Fig.~\ref{fig:main:circuit-for-C1} can be removed. We also employ mid-circuit measurements to realize multiple instances of $V_0$, reducing the required qubit count [see Appendix~\ref{section:appendix:circuits-for-Ag-and-C-F} for details]. These techniques together reduce the circuit depth to approximately one-tenth of a direct implementation: the median depth drops from 592.5 to 61 for $C_1$ and from 1056 to 95 for $C_2$, as shown in Appendix~\ref{section:appendix:depth-evaluation-corr}. The deeper circuits in this demonstration compared to the previous one are mainly due to the larger number of qubits and the operations required to measure $R[T]$. 

\subsubsection{Empirical results}
We conduct the demonstration by varying the parameters of $U_{SE}$ as well as the target observables $\mathcal{G}$ and $\mathcal{H}$ for $R[T]$. For each setting, we obtain expectation values by executing 4,000 measurements, repeated 20 times. The empirical results, shown in Fig.~\ref{fig:experiment-result-correlator}, provide the first empirical verification of thermodynamic trade-offs in quantum time correlators.

Figure~\ref{fig:experiment-result-correlator}a shows that the empirical thermodynamic inefficiency, given by ${\rm Var}[{\rm Re}[R(T)]]/({\rm Re}[R(T)]-\mathcal{C}_{\mathcal{F}_{\rm corr}})^2\times \mathcal{A}$,  exceeds unity, thereby confirming the QTUR for $R[T]$ stated in Eq.~\eqref{eq:generalQTUR-for-time-correlator}. Compared to the optimal observable $\mathcal{F}_{\rm opt}$ in Fig.~\ref{fig:experiment-result-Fopt}a, the inefficiency of ${\rm Re}[R(T)]$ deviates from the bound by roughly one order of magnitude. However, as the circuit in Fig.~\ref{fig:illustration}b was not optimized to reduce fluctuations, this small gap still suggests the practical relevance of $\mathcal{A}$ in designing circuits for measuring physical quantities. Further details of the empirical data are provided in Appendix~\ref{section:appendix:correlator-empirical-results}.

Figure~\ref{fig:experiment-result-correlator}b presents the trade-off results for ${\rm Re}[R(T)]$ as described by Eq.~\eqref{eq:timeCorrelationBound}. All empirical values of the inefficiency, quantified by $\sqrt{\mathcal{A}}/|{\rm Re}[R(T)]-\mathcal{C}_{\mathcal{G},\mathcal{H}}|$, satisfy the theoretical bound and approach the limit within a small constant factor, indicating the empirical tightness of the bound. Moreover, unlike $\mathcal{F}_{\rm opt}$, the inefficiency of ${\rm Re}[R(T)]$ shows no clear dependence on $\gamma$. In particular, for small $\gamma$, it falls below the theoretical prediction, which moderately increases as $\gamma$ decreases. This deviation is explained by the relatively large empirical values of $|{\rm Re}[R(T)]-\mathcal{C}_{\mathcal{G},\mathcal{H}}|$ observed in this regime. We provide a detailed analysis of empirical errors using numerical simulation in Appendix~\ref{section:appendix:correaltor-error-analysis}. Consequently, the primary source of the discrepancy is identified as depolarizing noise \cite{Nielsen2010-tb}, rather than thermal effects that dominated in the previous demonstration. These findings imply that the thermodynamic relation involving quantum time correlators is significantly influenced by open quantum dynamics arising from untraced environmental interactions.

\section{Discussion and conclusion}\label{section:conclusion}
We have established and empirically verified a general QTUR, which constrains the precision of \textit{arbitrary} observables under CPTP dynamics. The bound is governed solely by the survival activity $\mathcal{A}$ and remains tight: it is saturable by an optimal observable. Beyond precision bounds, the same framework yields trade-offs for quantum time correlators, thereby relating changes in correlations to the thermodynamic quantity $\mathcal{A}$.

The essential contribution of our demonstrations lies in bridging the gap between theory and practice on today's imperfect hardware. Real processors suffer from considerable gate errors, short coherence times, and state-preparation biases. As shown in Fig.~\ref{fig:experiment-result-compare_nonoptimized_and_eagle}, a straightforward approach fails to confirm the general QTUR. To overcome this, we developed a verification methodology suited to current devices: (i) a Neumann series expansion that avoids explicit matrix inversion, and (ii) circuit optimizations that exploit the theoretical structure of QTUR-related quantities. In combination, these techniques enable accurate QTUR verification on present-day processors with limited capabilities.

Using the optimal observable, we observed the first empirical saturation of the general QTUR, indicating that the bound is tight on real hardware, not just in idealized models. For quantum time correlators, the measured inefficiency approached the theoretical limit within a modest factor (ranging from several to several tens). Taken together, these results establish a fundamental QTUR that satisfies the three key criteria: (i) \textit{generality}, (ii) \textit{sharpness}, and (iii) \textit{empirical accessibility}. Hence, we conclude that the survival activity is a pivotal quantity dictating precision in quantum systems, and therefore,  measuring the survival activity alone tightly predicts achievable precision on the device. This insight could contribute to both the fundamental understanding of quantum information processing and the practical design of quantum technologies.

Our analysis of empirical errors then identifies the main factors behind the deviations from theoretical predictions: qubit temperature substantially relaxes the precision bound in the optimal-observable test, while residual open-system noise (notably depolarizing noise) alters the evolution of correlators. These findings not only highlight the physical implications accessible through our quantum-computer-based verification but also benchmark current hardware from a thermodynamic perspective.

This work extends the application of quantum computers in thermodynamic research \cite{Garcia-Perez2020-bb,Conlon2023-gw,Cech2023-jf} to the empirical demonstration of thermodynamic trade-off relations. Unlike earlier studies that employed quantum computers to verify fluctuation theorems~\cite{Gardas2018-xg, Solfanelli2021-rc, Zhang2022-yd}, which are formulated as equalities, our work investigates the empirical tightness of the QTUR. As discussed, accurate verification is crucial for assessing whether the inequality holds, particularly when testing near the theoretical limit. 

Viewing quantum computers as thermodynamic systems aligns with recent advances in quantum thermodynamics \cite{Blok2025-fh}. In particular, the finite temperature of qubits, which is inevitable due to the third law of thermodynamics, is of considerable interest: it limits the scalability of quantum computing \cite{Buffoni2022-jf} and has motivated the development of quantum heat engines for qubit cooling \cite{Aamir2025-qq}. Our findings, which highlight the significant effect of temperature on the general QTUR, complement this growing body of research. A promising direction for future work is thus to explore a QTUR for quantum heat engines implemented on quantum processors, as these engines are theoretically expected to obey the bound \cite{Miller2021-zn}. We expect our approach to inspire future exploration of the interplay between thermodynamic trade-off relations and quantum computing by establishing quantum computers as platforms for the empirical investigation of these relations.

\begin{acknowledgments}
    This work was supported by JSPS KAKENHI Grant Numbers JP23KJ0576 and JP23K24915. We acknowledge the use of IBM Quantum services for this work. The views expressed are those of the authors, and do not reflect the official policy or position of IBM or the IBM Quantum team.
\end{acknowledgments}

\appendix

\section{On the general QTUR [Eq.~\eqref{eq:GeneralTUR}]}

\subsection{Derivation}\label{section:appendix:derivation-of-generalQTUR}
In this section, we provide the full derivation of the general QTUR [Eq.~\eqref{eq:GeneralTUR}]. Generally, we assume that the initial state of the principal system \textit{S} denoted by $\rho_S(0)$ is a mixed state. The case of pure states is included in this general setting. Let us consider the composite system $R+S+E$, where $R$ is a virtual ancillary system used to purify the initial state of the principal system $S$. The initial state of the composite system $R+S+E$ is given by $\ket{\Psi_{RSE}(0)}=\ket{\Psi_{RS}(0)}\otimes\ket{0_E}$, where $\ket{\Psi_{RS}(0)}=\sum_i\sqrt{p_i}\ket{\psi_i}\otimes\ket{\psi_i}$. The reduced state on \textit{S} satisfies $\rho_S(0)={\rm Tr}_R[\ketbra{\Psi_{RS}(0)}]$. The final state of the composite system $R+S+E$ is given by $\ket{\Psi_{RSE}(T)}=(I_R\otimes U)(\ket{\Psi_{RS}(0)}\otimes\ket{0_E})=\sum_{im} \sqrt{p_i}\ket{\psi_i}\otimes V_m \ket{\psi_i}\otimes \ket{m_E}$ with Kraus operators $\{V_m\}$ of the CPTP map $\Phi$ for the principal system $S$. It should be noted that, if necessary, the first basis state of $E$ at $t=T$ may be chosen as a generic state $\ket{\phi_E^0}$. In this case we define $V_0=\bra{\phi_E^0}U_{SE}\ket{0_E}$ and proceed to the general QTUR.

As in the derivation of the pure state case in the main text [Section~\ref{section:brief-derivation-generalQTUR}], we use the quantum Cramér--Rao inequality [Eq.~\eqref{eq:end-matter-derivation-cramerrao}] to bound the variation of observables. The QFI is given as follows, similarly to Eq.~\eqref{eq:main-texti-QFI}:
\begin{align}
    J(\theta) &= 4\left[\ev**{H_1(\theta)}{\Psi_{RSE}(0)}\vphantom{\ev**{H_2(\theta)}{\Psi_{RSE}(0)}^2}\right.\nonumber\\
    &\quad\left.-\ev**{H_2(\theta)}{\Psi_{RSE}(0)}^2\right],\label{eq:QFI}
\end{align}
with
\begin{align}
H_1(\theta)&= I_R\otimes\sum_{m=0}^M \dv{V_m^\dagger(\theta)}{\theta}\dv{V_m}{\theta}\otimes I_{E},\\
H_2(\theta)&= I_R\otimes i\sum_{m=0}^M\dv{V_m^\dagger(\theta)}{\theta}V_m(\theta)\otimes I_E.
\end{align}
Note that the $i$ in the second equality represents the imaginary unit and should be distinguished from $i$ used as an index. The QFI $J(\theta=0)$ is calculated as $J(\theta)=\mathcal{A}$ with the survival activity $\mathcal{A}$ defined in Eq.~\eqref{eq:survial-activity} \cite{Hasegawa2021-aq}.

Next, we provide details of the calculation of $\partial_\theta \langle\mathcal{F}\rangle_\theta$ in Eq.~\eqref{eq:main-text-derivativeF}. $\expval{\mathcal{F}}_\theta$ is given by
\begin{widetext}
\begin{align}
    \expval{\mathcal{F}}_\theta
    &= \bra{\Psi_{RSE}(T)}\mathcal{F}\ket{\Psi_{RSE}(T)}\nonumber\\
    &= \qty(\sum_{ik} \sqrt{p_i}\bra{\psi_i}\otimes\bra{\psi_i}V_k(\theta)^\dagger\otimes \bra{k_E})\mathcal{F}\qty(\sum_{jl} \sqrt{p_j}\ket{\psi_j}\otimes V_l(\theta)\ket{\psi_j}\otimes\ket{l_E}).
\end{align}
The derivative of $\expval{\mathcal{F}}_\theta$ with respect to $\theta$ is given by
\begin{align}
    \partial_\theta \expval{\mathcal{F}}_\theta
    &= \qty(\sum_{ik} \sqrt{p_i}\bra{\psi_i}\otimes\bra{\psi_i}\dv{V_k(\theta)^\dagger}{\theta}\otimes \bra{k_E})\mathcal{F}\qty(\sum_{jl} \sqrt{p_j}\ket{\psi_j}\otimes V_l(\theta)\ket{\psi_j}\otimes\ket{l_E})\nonumber\\
    &\quad+ \qty(\sum_{ik} \sqrt{p_i}\bra{\psi_i}\otimes\bra{\psi_i}V_k(\theta)^\dagger\otimes \bra{k_E})\mathcal{F}\qty(\sum_{jl} \sqrt{p_j}\ket{\psi_j}\otimes \dv{V_l(\theta)}{\theta}\ket{\psi_j}\otimes\ket{l_E})\nonumber\\
    &= \qty(\sum_{ik} \sqrt{p_i}\bra{\psi_i}\otimes\bra{\psi_i}\dv{V_k(\theta)^\dagger}{\theta}\otimes \bra{k_E})\mathcal{F}\qty(\sum_{jl} \sqrt{p_j}\ket{\psi_j}\otimes V_l(\theta)\ket{\psi_j}\otimes\ket{l_E})+{\rm h.c.}
\end{align}
\end{widetext}
For $k>0$, the derivative of $V_k(\theta)$ is given by
\begin{align}
    \dv{V_k(\theta)}{\theta} &= \frac{1}{2}e^{\theta/2}V_k(\theta)\eval{=}_{\theta=0} \frac{1}{2}V_k.
\end{align}
For $k=0$, we need some calculations to obtain the derivative. First, we introduce the spectral decomposition of $\sum_{m>0}^M V_m^\dagger V_m$ as
\begin{align}
    \sum_{m>0}^M V_m^\dagger V_m = \sum_n \zeta_n \Pi_n,
\end{align}
where $\zeta_n$ is the eigenvalue of $\sum_{m>0}^M V_m^\dagger V_m$ and $\Pi_n$ is the projection operator onto the eigenspace of $\zeta_n$. Then, we can write $V_0(\theta)$ as
\begin{align}
    V_0(\theta) &= U_0\sum_n \sqrt{1-e^{\theta}\zeta_n}\Pi_n.
\end{align}
Using this equation, we can calculate the derivative $\dv{V_0(\theta)}{\theta}$ as
\begin{align}
    \dv{V_0(\theta)}{\theta} &= \frac{1}{2}U_0\sum_n \frac{-\zeta_n e^{\theta/2}}{\sqrt{1-e^{\theta}\zeta_n}}\Pi_n\nonumber\\
    &\eval{=}_{\theta=0} \frac{1}{2}U_0\sum_n \frac{-\zeta_n}{\sqrt{1-\zeta_n}}\Pi_n\nonumber\\
    &= \frac{1}{2}U_0\sum_n \sqrt{1-\zeta_n}\Pi_n - \frac{1}{2}U_0\sum_n \frac{1}{\sqrt{1-\zeta_n}}\Pi_n\nonumber\\
    &= \frac{1}{2}V_0 - \frac{1}{2}V_0^{\dagger^{-1}}.
\end{align}
Thus we obtain the derivative of $\expval{\mathcal{F}}_\theta$ as
\begin{widetext}
\begin{align}
    \partial_\theta \expval{\mathcal{F}}_{\theta=0}
    &= \frac{1}{2} \qty(\sum_{ik} \sqrt{p_i}\bra{\psi_i}\otimes\bra{\psi_i}V_k^\dagger\otimes \bra{k_E})\mathcal{F}\qty(\sum_{jl} \sqrt{p_j}\ket{\psi_j}\otimes V_l\ket{\psi_j}\otimes\ket{l_E})\nonumber\\
    &\quad- \frac{1}{2} \qty(\sum_{i} \sqrt{p_i}\bra{\psi_i}\otimes\bra{\psi_i}V_0^{-1} \otimes \bra{0_E})\mathcal{F}\qty(\sum_{jl} \sqrt{p_j}\ket{\psi_j}\otimes V_l\ket{\psi_j}\otimes\ket{l_E})\nonumber\\
    &\quad+({\rm h.c.\; of\; the\; first\; and\; second\; terms})\nonumber\\
    &=\qty(\sum_{ik} \sqrt{p_i}\bra{\psi_i}\otimes\bra{\psi_i}V_k^\dagger\otimes \bra{k_E})\mathcal{F}\qty(\sum_{jl} \sqrt{p_j}\ket{\psi_j}\otimes V_l\ket{\psi_j}\otimes\ket{l_E})\nonumber\\
    &\quad- \frac{1}{2} \qty(\sum_{i} \sqrt{p_i}\bra{\psi_i}\otimes\bra{\psi_i}V_0^{-1} \otimes \bra{0_E})\mathcal{F}\qty(\sum_{jl} \sqrt{p_j}\ket{\psi_j}\otimes V_l\ket{\psi_j}\otimes\ket{l_E})\nonumber\\
    &\quad- \frac{1}{2} \qty(\sum_{ik} \sqrt{p_i}\bra{\psi_i}\otimes\bra{\psi_i}V_k^\dagger\otimes \bra{k_E})\mathcal{F}\qty(\sum_{j} \sqrt{p_j}\ket{\psi_j}\otimes V_0^{\dagger^{-1}}\ket{\psi_j}\otimes\ket{0_E})\nonumber\\
    &=\expval{\mathcal{F}} - \frac{1}{2} \bra*{\widetilde{\Psi}_{RSE}(0)}\mathcal{F}\ket{\Psi_{RSE}(T)} - \frac{1}{2} 
    \bra{\Psi_{RSE}(T)}\mathcal{F}\ket*{\widetilde{\Psi}_{RSE}(0)}\nonumber\\
    &= \expval{\mathcal{F}} - \mathcal{C}_\mathcal{F},\label{eq:appendix:scaling}
\end{align}
\end{widetext}
where $\ket*{\widetilde{\Psi}_{RSE}(0)}=(I_R\otimes {V_0^\dagger}^{-1}\otimes I_E)\ket{\Psi_{RS}(0)}\otimes\ket{0_E}$ and $\mathcal{C}_\mathcal{F}={\rm Re}[\bra*{\widetilde{\Psi}_{RSE}(0)}\mathcal{F}\ket*{\Psi_{RSE}(T)}]$. Combining Eqs.~\eqref{eq:survial-activity}, \eqref{eq:end-matter-derivation-cramerrao}, and \eqref{eq:appendix:scaling}, we obtain the general QTUR [Eq.~\eqref{eq:GeneralTUR}] for $R$+$S$+$E$.

Several points should be noted regarding the above derivation. First, it holds for a generic observable $\mathcal{F}$ on the composite system $R+S+E$. In practice, we may choose $\mathcal{F}$ to be an observable solely on the system $S+E$, as the ancillary system $R$ is virtual. Second, when the initial state of $S$ is pure, we substitute $\ket{\Psi_{RS}(0)}=\ket{\Psi_S(0)}\otimes\ket{\Psi_S(0)}$ into the derivation. Third, the equality condition of the general QTUR comes from that of the quantum Cramér--Rao inequality in Eq.~\eqref{eq:end-matter-derivation-cramerrao}. The equality in Eq.~\eqref{eq:end-matter-derivation-cramerrao} is attained when the observable $\mathcal{F}$ is proportional to the symmetric logarithmic derivative $\mathcal{L}$ with respect to the Kraus operators and the perturbation parameter $\theta$ \cite{Hotta2004-px}. $\mathcal{L}$ is given by
\begin{align}
    \mathcal{L}&=\ketbra{\Psi_{SE}(T)}\nonumber\\
    &\quad-\frac{1}{2}(\ketbra*{\widetilde{\Psi}_{SE}(0)}{\Psi_{SE}(T)}+\ketbra*{\Psi_{SE}(T)}{\widetilde{\Psi}_{SE}(0)}).\label{eq:appendix:logarithmic-derivative-L}
\end{align}
Consequently, the equality condition of the general QTUR is satisfied when $\mathcal{F}$ is proportional to $\mathcal{L}$. This condition can be fulfilled by considering observables solely on $S+E$ (excluding $R$) when the initial state of $S$ is pure.

\subsection{Example: The $U_{SE}$ from our demonstrations}\label{section:appendix:derivation-U-SE-example}
\begin{figure}[h]
    \centering
    \includegraphics[width=0.7\linewidth]{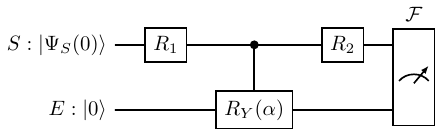}
    \caption{Parameterized circuit implementing $U_{SE}$ using a controlled-RY gate.}
    \label{fig:end-matter-circuit}
\end{figure}
In this section, we illustrate our theory using a concrete example system from our demonstrations. The core component realizing the system-environment coupling $U_{SE}$ is a controlled-RY gate, depicted in Fig.~\ref{fig:end-matter-circuit}. For simplicity, we consider a two-qubit system where the system qubit $S$ undergoes single-qubit rotations $R_1$ and $R_2$. The unitary operator $U_{SE}$ for this circuit is given by
\begin{align}
    U_{SE} = &\qty(R_2\otimes I_E)\times\\\nonumber
    &\qty(\ketbra{0}{0}\otimes I_E+ \ketbra{1}{1}\otimes R_Y(\alpha))\qty(R_1\otimes I_E),
\end{align}
where $\alpha$ is the rotation angle of the RY gate acting on the environment qubit $E$. The RY gate is defined as
\begin{align}
    R_Y(\alpha)=
    \begin{pmatrix}
        \cos\frac{\alpha}{2} & -\sin\frac{\alpha}{2}\\
        \sin\frac{\alpha}{2} & \cos\frac{\alpha}{2}
    \end{pmatrix}.
\end{align}

In the main text, $\alpha$ is related to the interaction strength $\gamma$ via $\alpha=\pi\gamma$. The corresponding primary Kraus operator $V_0 = \bra{0_E}U_{SE} \ket{0_E}$ is
\begin{align}
    V_0 = R_2
    \begin{pmatrix}
        1 & 0 \\
        0 & \bra{0} R_Y(\alpha) \ket{0}
    \end{pmatrix}
    R_1. \label{eq:end-matter-V0}
\end{align}
Note that $\bra{0} R_Y(\alpha) \ket{0} = \cos\frac{\alpha}{2}$. 

By definition, $V_0$ yields the (unnormalized) state of system $S$ when a measurement on the environment $E$ yields the outcome $\ket{0}$ after the interaction $U_{SE}$. If $\alpha=0$ (no interaction), $V_0$ simplifies to the unitary operator $R_2R_1$, and the Kraus operator corresponding to outcome $\ket{1}$ vanishes.

Consider the weak-coupling regime where $\alpha$ is small. In this regime, we perform a perturbative expansion in $\alpha$, while keeping the parameter $\theta$ fixed. As a result of the $\alpha$-perturbation, the $\theta$-perturbed Kraus operator $V_0(\theta)$ from Eq.~\eqref{eq:end-matter-perturbation-V0} becomes approximately
\begin{align}
    V_0(\theta)\approx R_2
    \begin{pmatrix}
        1 &0\\0&\bra{0}R_Y(\alpha e^{\theta/2})\ket{0}
    \end{pmatrix}
    R_1,
\end{align}

Therefore, the virtual perturbation effectively scales the rotation angle of the controlled-RY gate to $\alpha e^{\theta/2}$, thereby modifying the system-environment interaction strength.

Next, we examine the survival activity $\mathcal{A}$. The operator ${V_0}^\dagger V_0$ is given by
\begin{align}
    {V_0}^\dagger V_0 = {R_1}^\dagger \begin{pmatrix}1 &0\\0&|\bra{0}R_Y(\alpha)\ket{0}|^2\end{pmatrix}R_1.
\end{align}
Its expectation value with respect to the initial system state $\rho_S(0)$ represents the probability of the environment $E$ remaining in state $\ket{0}$ after the interaction. The survival activity $\mathcal{A}$ is related to the expectation value of the inverse operator $(V_0^\dagger V_0)^{-1}$:
\begin{align}
    ({V_0}^\dagger V_0)^{-1} = {R_1}^\dagger \begin{pmatrix}1 &0\\0&\frac{1}{|\bra{0}R_Y(\alpha)\ket{0}|^2}\end{pmatrix}R_1,
\end{align}
and consequently
\begin{align}
    \mathcal{A}\approx \frac{\alpha^2}{4} \expval{{R_1}^{\dagger}\ket{1}\bra{1}R_1}_{\rho_S(0)},
\end{align}
when $\alpha\ll 1$. Thus, we conclude that for small $\gamma$:
\begin{align}
    \mathcal{A}\propto \gamma^2.
\end{align}

Finally, we consider the inherent-dynamics contribution $\mathcal{C}_\mathcal{F}$. For simplicity, we analyze a separable observable $\mathcal{F}^{\rm sep}$, although the optimal observable $\mathcal{F}_{\rm opt}$ discussed in the main text is generally non-separable (global). A separable observable can be written as $\mathcal{F}^{\rm sep}=\sum_{l,m}f_{lm}\ketbra{\phi_l}{\phi_l} \otimes \ketbra{m_E}{m_E}$, where $\{\ket{\phi_l}\}$ is an eigenbasis for system $S$. In this case, $\mathcal{C}_\mathcal{F}$ simplifies to
\begin{align}
    \mathcal{C}_{\mathcal{F}^{\rm sep}}
    &= {\rm Re}\qty[{\rm Tr}[\rho_{S}(0)({V_0}^{-1}\mathcal{F}^{\rm sep}_0V_0)]],\label{eq:end-matter-QGsep-intermediate}
\end{align}
where $\mathcal{F}_0^{\text{sep}} = \sum_{l} f_{l0}\ketbra{\phi_l}{\phi_l}$ is the part of the observable $\mathcal{F}^{\text{sep}}$ conditioned on the environment $E$ being in state $\ket{0}$.

Substituting the expression for $V_0$ from Eq.~\eqref{eq:end-matter-V0} into Eq.~\eqref{eq:end-matter-QGsep-intermediate} and expanding for small $\alpha \ll 1$, we obtain
\begin{align}
    &\mathcal{C}_{\mathcal{F}_{\rm sep}}\approx {\rm Tr}[R_2R_1 \rho_S(0) R_1^\dagger R_2^\dagger\mathcal{F}_0^{\rm sep}]+\nonumber\\
    &\quad\frac{\alpha^2}{8}{\rm Re}\left[{\rm Tr}\left[R_1 \rho_S(0)R_1^\dagger\left[
    \ket{1}\bra{1}
    , R_2^\dagger \mathcal{F}_0^{\rm sep} R_2
    \right]\right]\right]+\mathcal{O}(\alpha^4),\label{eq:end-matter-CF-epsilon}
\end{align}
where $[A,B]=AB-BA$ is the commutator.

The leading term in Eq.~\eqref{eq:end-matter-CF-epsilon} corresponds to the expectation value of $\mathcal{F}_0^{\rm sep}$ evolved solely by the system unitary $R_2 R_1$, i.e., the value in the absence of system-environment interaction ($\alpha=0$). Hence, the quantity $(\expval{\mathcal{F}}-\mathcal{C}_\mathcal{F})^2$ appearing in the general QTUR [Eq.~\eqref{eq:GeneralTUR}] represents the contribution to the observable $\mathcal{F}$ driven solely by the system-environment interaction. The second term, proportional to $\alpha^2$, involves the expectation of a commutator evaluated for the state $R_1 \rho_S(0) R_1^\dagger$. This term captures quantum effects: for instance, if $R_1 \rho_S(0) R_1^\dagger$ and $R_2^\dagger \mathcal{F}_0^{\rm sep} R_2$ are both diagonal in the computational basis, the commutator term vanishes. Therefore, the non-zero contribution of this term in our demonstrations indicates the presence of quantum coherence effects.

\subsection{Generalization to thermal environments}\label{section:appendix:generalization-to-thermal-environments}
The aforementioned derivation of the general QTUR assumes the environment is in a pure state. While such pure-state environments can model quantum jump records as in Ref.~\cite{Hasegawa2021-aq}, real environments exhibit finite temperatures, resulting in thermal states. Here, we extend our framework to encompass generic thermal environments.

Consider an environment $E$ initially in a thermal state $\rho_E^{(\rm{th})}=\sum_i \lambda_i \ketbra{\phi_i}$, where $\{\ket{\phi_i}\}$ forms an energy eigenbasis with $\ket{\phi_0}$ being the ground state. The population $\lambda_i$ is given by $\lambda_i=e^{-\beta E_i}/Z$, where $\beta$ is the inverse temperature of the environment, $E_i$ is the eigenenergy corresponding to $\ket{\phi_i}$, and $Z$ is the partition function. For this thermal state, the Kraus operators $\{V_{jk}\}$ associated with the unitary operator $U_{SE}$ on the composite system $S$+$E$ are defined as
\begin{align}
    V_{jk} = \sqrt{\lambda_j}\bra{\phi_k}U_{SE}\ket{\phi_j},
\end{align}
which satisfies the completeness condition $\sum_{jk}V_{jk}^\dagger V_{jk}=I$. The final state of the principal system $S$ is then expressed as $\rho_S(T)=\sum_{jk}V_{jk}\rho_S(0)V_{jk}^\dagger$.

To derive a modified general QTUR for thermal environments, we extend the derivation presented in the previous section using these new Kraus operators. We can show that the quantum Cramér--Rao inequality in Eq.~\eqref{eq:end-matter-derivation-cramerrao} remains valid for the QFI $J(\theta)$ in Eq.~\eqref{eq:main-texti-QFI} with these new Kraus operators $\{V_{jk}\}$. We introduce a special Kraus operator $\overline{V_0}$ corresponding to the case where the environment remains in its ground state $\ket{\phi_0}$, i.e.,
\begin{align}
    \overline{V_0}\coloneq V_{00}.\label{reply:eq:V0-def}
\end{align}
With this definition, the new QFI equals the modified survival activity:
\begin{align}
    \overline{\mathcal{A}}={\rm Tr}[\rho_S(0)(\overline{{V_0}}^\dagger \overline{V_0})^{-1}]-1.
\end{align}
Next, we consider the modified inherent-dynamics contribution $\overline{\mathcal{C}_\mathcal{F}}$. Following Eq.~\eqref{eq:appendix:scaling}, we obtain
\begin{align}
    \overline{\mathcal{C}_\mathcal{F}}=\sqrt{\lambda_0}{\rm Re}[{\rm Tr}[(\rho_S(0)\otimes \ketbra*{\phi_0})({\overline{V_0}}^{-1}\otimes I_E)\mathcal{F}U]].
\end{align}
These quantities lead to the general QTUR for thermal environments:
\begin{equation}
    \frac{{\rm Var}[\mathcal{F}]}{(\expval{\mathcal{F}}-\overline{\mathcal{C}_\mathcal{F}})^2}\geq\frac{1}{\overline{\mathcal{A}}}.
\end{equation}
This relation generalizes the original theory to finite-temperature environments. In the limit of zero temperature, the modified general QTUR reduces to the original relation.

\subsection{Simplification of $\mathcal{C}_{\mathcal{F}}$ for specific observables}\label{section:appendix:simplification-of-CF-for-Fs}
For a separable observable $\mathcal{F}^{\rm sep}$ on the composite system $S+E$, the inherent-dynamics contribution $\mathcal{C}_\mathcal{F}$ in the general QTUR [Eq.~\eqref{eq:GeneralTUR}] takes a simplified form. $\mathcal{F}^{\rm sep}$ can be written as  $\mathcal{F}^{\rm sep}=\sum_{l,m}f_{lm}\ketbra*{\phi^{(\rm S)}_l}\otimes\ketbra*{\phi^{(E)}_m}$, with  $\ket*{\phi^{(\rm S)}_l}$ and $\ket*{\phi^{(E)}_m}$ being the eigenbases of subsystems \textit{S} and \textit{E}. We define the Kraus operators $\{V_m\}$ by the basis $\ket*{\phi^{(E)}_m}$.  In this case, $\mathcal{C}_\mathcal{F}$ reduces to
\begin{align}
    \mathcal{C}_{\mathcal{F}^{\rm sep}}
    &= {\rm Re}\qty[\bra{\Psi_{\rm SE}(0)}(I_{\rm R}\otimes V_0^{-1}\otimes I_{E})\mathcal{F}^{\rm sep}\ket{\Psi_{\rm SE}(T)}]\nonumber\\
    &= {\rm Re}\qty[{\rm Tr}[\rho_{S}(0)({V_0}^{-1}\mathcal{F}^{\rm sep}_0V_0)]],\label{eq:appendix:QGsep-intermediate}
\end{align}
where $\mathcal{F}_0^{\text{sep}} = \sum_{k} f_{k0}|\phi_k^{(S)}\rangle\langle\phi_k^{(S)}|$ is the observable $\mathcal{F}^{\text{sep}}$ conditioned on the environment \textit{E} being in the state $|\phi^{(E)}_0\rangle$. For local observables $\mathcal{F}_S$ acting solely on subsystem $S$, a further simplification yields
\begin{align}
    \mathcal{C}_{{\mathcal{F}_S}}= {\rm Re}\qty[{\rm Tr}[\rho_{S}(0)({V_0}^{-1}\mathcal{F}_S V_0)]].\label{eq:appendix:C-Fs}
\end{align}
We leverage this expression in the circuit optimization of the second demonstration [Section~\ref{section:demonstration-II-circuit-optimization}].

For local observables $\mathcal{F}_{E}=\sum_m f_m \ketbra*{\phi_m^{(E)}}$ on $E$, such as the number of emitted photons,  $\mathcal{C}_{\mathcal{F}_{E}}$ becomes a constant as $\mathcal{C}_{\mathcal{F}_{E}}=f_0$. In particular, when $\mathcal{F}_{E}$ is a counting observable, which satisfies $f_0=0$, $\mathcal{C}_{\mathcal{F}_{E}}$ vanishes and the general QTUR reduces to the previous QTUR \cite{Hasegawa2021-aq}.

\subsection{Inherent-dynamics contribution $\mathcal{C}_\mathcal{F}$ in the weak coupling regime}\label{section:appendix:CF-weak-coupling}
In this section, we consider the weak coupling regime between a principal system $S$ and an environment $E$, where the dynamics on $S$ is nearly unitary. In this case, the Kraus operator $V_0$ is almost unitary and satisfies $V_0^\dagger V_0=I-\epsilon$, where $\epsilon$ is an operator with a small norm. Using polar decomposition, $V_0$ can be expressed in terms of $\epsilon$ as $V_0=U_0\sqrt{V_0^\dagger V_0}=U_0\sqrt{I-\epsilon}$, where $U_0$ is a unitary operator. We can approximate $V_0$ as $V_0\approx U_0(I-\frac{1}{2}\epsilon)$, and its inverse as $V_0^{-1}\approx (I+\frac{1}{2}\epsilon)U_0^\dagger$. Using this approximation, $\mathcal{C}_\mathcal{F}$ can be approximated as
\begin{align}
    \mathcal{C}_\mathcal{F}
    &\approx {\rm Re}[\bra{\Psi_S(0)}(I+\frac{1}{2}\epsilon)U_0^\dagger\otimes\bra{0_E}\mathcal{F}U_{SE}\ket{\Psi_{SE}(0)}]\nonumber\\
    &={\rm Re}[\bra{\Psi_S(0)}U_0^\dagger\otimes\bra{0_E}\mathcal{F}U_{SE}\ket{\Psi_{SE}(0)}]\nonumber\\
    &\quad+\frac{1}{2}{\rm Re}[\bra{\Psi_S(0)}\epsilon U_0^\dagger\otimes\bra{0_E}\mathcal{F}U_{SE}\ket{\Psi_{SE}(0)}]
\end{align}
For local observables on $S$, where $\mathcal{C}_\mathcal{F}$ simplifies to Eq.~\eqref{eq:appendix:C-Fs}, we obtain
\begin{align}
    &\mathcal{C}_{\mathcal{F}_S}
    \approx {\rm Re}[\bra{\Psi_S(0)}(I-\frac{1}{2}\epsilon+\epsilon)U_0^\dagger \mathcal{F}_S V_0\ket{\Psi_S(0)}]\nonumber\\
    &=\bra{\Psi_S(0)}V_0^\dagger\mathcal{F}_S V_0\ket{\Psi_S(0)}+{\rm Re}[\bra{\Psi_S(0)}\epsilon U_0^\dagger \mathcal{F}_S V_0\ket{\Psi_S(0)}]\nonumber\\
    &\approx\bra{\Psi_S(0)}V_0^\dagger\mathcal{F}_S V_0\ket{\Psi_S(0)}+{\rm Re}[\bra{\Psi_S(0)}\epsilon U_0^\dagger \mathcal{F}_S U_0\ket{\Psi_S(0)}].
\end{align}
Thus, the primary term of $\mathcal{C}_{\mathcal{F}_S}$ is the expectation value of ${\mathcal{F}_S}$ under the non-trace-preserving dynamics described by $V_0$.

\subsection{Measurement of the survival activity $\mathcal{A}$}\label{section:appendix:measurement-of-A-theory}
In this section, we present a methodology to quantify the survival activity $\mathcal{A}$ using higher-order approximations beyond those considered in the main text [Section \ref{section:main:measure-A-and-CF}]. We begin by assuming the feasibility of applying the inverse unitary operation, $U_{SE}^\dagger$, on the combined system and environment \textit{S}+\textit{E}, and that both the initial and final states on \textit{E} defining $V_0$ are given by $\ket{0_E}$. Later, we discuss the scenario where the reversibility of the unitary transformation associated with the CPTP map on \textit{S} is not available.

Consider truncating the series expansion of $\mathcal{A}$ in Eq.~\eqref{eq:survial-activity} by limiting the summation over $n$ to $N$. We present a protocol to measure ${\rm Tr}[\rho_{S}(0)(V_0^\dagger V_0)^{N}]$, which allows for the simultaneous measurement of ${\rm Tr}[\rho_{S}(0)(V_0^\dagger V_0)^{n}]$ for all $n<N$. We define an unnormalized state $\rho_{S}^{(n)}$ that satisfies ${\rm Tr}[\rho_{S}^{(n)}]={\rm Tr}[\rho_{S}(0)(V_0^\dagger V_0)^{n}]$. The initial state $\rho_{S}^{(0)}$ is given by $\rho_{S}^{(0)}=\rho_{S}(0)$. For $n>0$, the state $\rho_{S}^{(n)}$ is iteratively constructed as
\begin{align}
    \rho_{S}^{(n+1)} = 
    \begin{cases}
        V_0\rho_{S}^{(n)}V_0^\dagger & \text{if $n$ is even}, \\
        V_0^\dagger\rho_{S}^{(n)}V_0 & \text{if $n$ is odd}.\label{eq:appendix:rho-iteration}
    \end{cases}
\end{align}
Here, we exploit the cyclic property of the trace to obtain these relationships. For even $n$, $\rho_{S}^{(n+1)}$ is obtained by evolving $\rho_{S}^{(n)}$ under the operator $V_0$: first apply $U_{SE}$ to $\rho_{S}^{(n)} \otimes \ketbra{0_E}$, and then project onto the environmental state $\ket{0_E}$. For odd $n$, a similar process is employed using $V_0^\dagger$. The evolution with $V_0^\dagger$ follows from the definition of $V_0$, given by $V_0 = (I_{S} \otimes \bra{0_E})U_{SE}(I_{S} \otimes \ket{0_E})$. Consequently, $V_0^\dagger$ is defined as $V_0^\dagger = (I_{S} \otimes \bra{0_E})U_{SE}^\dagger(I_{S} \otimes \ket{0_E})$, which can be realized by first applying $U_{SE}^\dagger$ to $\rho_{S}^{(n)} \otimes \ketbra{0_E}$, followed by a projection onto the environmental state $\ket{0_E}$.

To evaluate ${\rm Tr}[\rho_{S}(0)(V_0^\dagger V_0)^{N}]$, we initialize the principal system $S$ in the state $\rho_{S}(0)$ and prepare $N$ replicas of the environmental system $E$ in the state $\ket{0_E}$. We then apply either $U_{SE}$ or $U_{SE}^\dagger$ to $S$ and each environmental replica ${E}^{(n)}$ according to the sequence prescribed by Eq.~\eqref{eq:appendix:rho-iteration}. After performing this sequence of operations $N$ times, we measure the environmental systems in the basis $\{\ket{0_E},\ket{1_E}\}$ and record the instances where all environments are found in the state $\ket{0_E}$. By repeating this procedure multiple times, we obtain an estimate of the probability of such occurrences, from which we can infer the desired expectation value ${\rm Tr}[\rho_{S}(0)(V_0^\dagger V_0)^{N}]$. Importantly, this approach allows us to simultaneously obtain the expectations ${\rm Tr}[\rho_{S}(0)(V_0^\dagger V_0)^{n}]$ for all $n<N$ using the same outcomes, simply by considering only the first $n$ environmental systems in each instance. Finally, the survival probability can be calculated using Eq.~\eqref{eq:appendix:xi-expansion} based on these expectation values.

We provide two additional insights regarding the protocol. First, the protocol could be implemented using only a single environmental system by incorporating measurements within each iteration. After applying $U_{SE}$ or $U_{SE}^\dagger$, we measure the environment in the $\{\ket{0_E},\ket{1_E}\}$ basis. If the outcome is $\ket{0_E}$, we proceed to the next iteration; if the outcome is $\ket{1_E}$, we discard the result and restart the protocol from the beginning. By keeping track of the number of successful iterations, we can estimate the expectation values ${\rm Tr}[\rho_{S}(0)(V_0^\dagger V_0)^{n}]$ for all $n\leq N$. Second, the survival activity $\mathcal{A}$ can be approximated even when $U_{SE}^\dagger$ is unavailable. By truncating the expansion in Eq.~\eqref{eq:appendix:xi-expansion} at $N=1$, we obtain the approximation $\mathcal{A} \approx 1  -  {\rm Tr}[\rho_{S}(0)V_0^\dagger V_0]$. This approximation is valid in the regime where the interaction between systems \textit{S} and \textit{E} is weak, as higher-order terms in the expansion become negligible in this case.

\subsection{Measurement of the inherent-dynamics contribution $\mathcal{C}_\mathcal{F}$\label{section:appendix:C-F}}
The inherent-dynamics contribution $\mathcal{C}_\mathcal{F}$ can be measured using an approximation method similar to that used for the survival activity $\mathcal{A}$. Recall the definition of $\mathcal{C}_\mathcal{F}$: $\mathcal{C_F}={\rm Re}[\mel*{\widetilde{\Psi}_{SE}(0)}{\mathcal{F}}{\Psi_{SE}(T)}]$ with the unnormalized state $\ket*{\widetilde{\Psi}_{SE}(0)}=({V_0^{-1}})^{\dagger}\ket{\Psi_{S}(0)}\otimes\ket{0_E}$. We rewrite $\mathcal{C}_\mathcal{F}$ as
\begin{align}
    \mathcal{C_F}={\rm Re}[\bra{\Psi_S(0)}\otimes\bra{0_E}({V_0}^{-1}\otimes I_E) \mathcal{F}U_{SE}\ket{\Psi_{S}(0)}\otimes \ket{0_E}].    
\end{align}
We further calculate it to obtain the expression $(V_0 V_0 ^\dagger)^{-1}$ as
\begin{align}
    \mathcal{C_F}&={\rm Re}[\bra{\Psi_S(0)}V_0^\dagger\otimes\bra{0_E}({(V_0V_0^\dagger)}^{-1}\otimes I_E)\nonumber\\
    &\quad\times\mathcal{F}U_{SE}\ket{\Psi_{S}(0)}\otimes \ket{0_E}].    
\end{align}
By the Neumann series expansion as in the previous section for $\mathcal{A}$, we can approximate $(V_0 V_0 ^\dagger)^{-1}$ as $(V_0 V_0 ^\dagger)^{-1}\approx 2I-V_0V_0^\dagger$ by truncating at $N=1$ in Eq.~\eqref{eq:appendix:V0-expansion}. Subsequently, we obtain the following first-order approximation:
\begin{align}
    \mathcal{C_F}
    \approx 2C_1
    -C_2,\label{eq:appendix:CF-approx}
\end{align}
where
\begin{align}
    C_1&={\rm Re}[(\bra{\Psi_S(0)}V_0^\dagger\otimes\bra{0_E}) \mathcal{F}U_{SE}(\ket{\Psi_{S}(0)}\otimes \ket{0_E})],\label{eq:appendix:C-1-def}\\
    C_2&={\rm Re}[\bra{\Psi_S(0)}V_0^\dagger\otimes\bra{0_E}(V_0V_0^\dagger\otimes I_E)\nonumber\\
    &\quad\times\mathcal{F}U_{SE}\ket{\Psi_{S}(0)}\otimes \ket{0_E}].\label{eq:appendix:C-2-def}
\end{align}

To measure the correlators $C_1$ and $C_2$, we employ the Hadamard test \cite{Somma2002-po}. Let us illustrate the test with a simple example of measuring $\mathrm{Re}\left[\bra{\Psi(0)}V^\dagger U\ket{\Psi(0)}\right]$, as shown in Fig.~\ref{fig:main:circuit-for-C1}a. In this test, we introduce an ancilla qubit $S'$ prepared in the state $\ket{+}$. The ancilla is then entangled with the system $S$ using a controlled-$U$ operation, which applies the unitary operator $U$ to the system only when the control qubit $S'$ is in the state $\ket{1}$. Next, a controlled-$V$ operation is applied, where the control is denoted by a white circle in the figure. This operation applies the unitary operator $V$ to the system $S$ only when the ancilla qubit $S'$ is in the state $\ket{0}$. Finally, to obtain $\mathrm{Re}\left[\bra{\Psi(0)}V^\dagger U\ket{\Psi(0)}\right]$, we measure the expectation value of the Pauli operator $\sigma_x$ on the ancilla qubit $S'$. In practice, this measurement can be realized by applying an Hadamard gate followed by a $\sigma_z$ measurement on $S'$. The Hadamard test provides a systematic approach to measure correlators by applying controlled unitary operations. By extending this method, we can measure the correlators $C_1$ and $C_2$ in our study.

\begin{figure}[ht]
    \centering
    \includegraphics[width=0.45\textwidth]{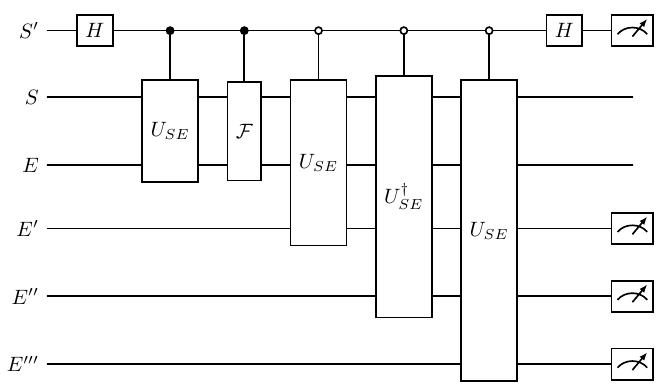}
    \caption[Straightforward implementation of the Hadamard tests for $\mathcal{C}_\mathcal{F}$]{\label{fig:appendix:circuits-for-C2-naive}Straightforward implementation of the circuits for the Hadamard tests to measure $\mathcal{C}_2$. All measurements are made in the computational basis $\{\ket{0},\ket{1}\}$. The white control represents applying the target unitary operator when the control qubit is in the state $\ket{0}$.
    }
\end{figure}

The straightforward implementations of the Hadamard tests for $C_1$ and $C_2$ are illustrated in Figs.~\ref{fig:main:circuit-for-C1}b and \ref{fig:appendix:circuits-for-C2-naive}, respectively. Here, if $\mathcal{F}$ is not unitary, we employ the unitary operation $\mathcal{F}_i$ satisfying $\sum_i \mathcal{F}_i=\mathcal{F}$ instead of $\mathcal{F}$. However, such straightforward implementations are impractical for real processors due to their excessive depths. Controlled unitary operations involving three or more qubits are generally challenging to implement when only single or two-qubit gates on sparsely connected qubits are available on real hardware. To address the issue of circuit depth, we develop shallow circuits to measure $C_1$ and $C_2$ as described in the main text. See Appendices~\ref{section:appendix:circuits-for-C-F-opt} and \ref{section:appendix:circuits-for-Ag-and-C-F} for details of optimization.

\section{IBM's quantum processors}\label{section:appendix:IBM}
\begin{table*}[ht]
\center
\caption{Specifications of IBM's quantum processors}
\label{table:processor_specifications}
\begin{tabular}{lcc}
\toprule

 & ibm\_torino & ibm\_sherbrooke \\
 \midrule
Two-qubit (CZ or ECR) gate error & $(4.56 \pm 0.19) \times 10^{-3}$ & $(7.75 \pm 0.43) \times 10^{-3}$\\
SX gate error & $(3.15 \pm 0.15) \times 10^{-4}$ & $(2.31 \pm 0.13) \times 10^{-4}$\\
Readout error & $(1.95 \pm 0.10) \times 10^{-2}$ & $(1.14 \pm 0.09) \times 10^{-2}$\\
T1 time & $(167.73 \pm 6.14)\mu\text{s}$ & $(268.19 \pm 7.96)\mu\text{s}$\\
T2 time & $(130.19 \pm 5.90)\mu\text{s}$& $(182.58 \pm 7.35)\mu\text{s}$\\
\bottomrule
\end{tabular}
\end{table*}
For our demonstrations, we use a superconducting qubit system from IBM's ``Heron'' quantum processor (ibm\_torino) \cite{Ibm_undated-ip,Ibm_undated-yk}. The processor comprises 133 transmon qubits \cite{Koch2007-qf}, each connected to two or three neighboring qubits, and all initialized in the state $\ket{0}$. The system supports five native gates: Identity, X, SX (the square root of X), RZ, and the two-qubit CZ (controlled Z) gate. Other gates are built from these primitive gates. Although we describe operations using non-native gates like RX and RY in the following, the circuits containing these gates are transpiled before execution on the quantum processor. Since only the measurement of $\sigma_z$ is available for the qubits, an Hadamard gate and $\sigma_z$ measurement are used to measure $\sigma_x$, for example. 

The qubits and gates on the ibm\_torino processor were calibrated daily during our demonstrations in May 2024. The processor's typical metrics over this period are shown in Table~\ref{table:processor_specifications}. For each metric, the median value across all qubits or gates was calculated for each day. The mean and standard deviation of these daily medians were then computed for the month, considering only days with available data from Ref.~\cite{Ibm_undated-ip}. The metrics for CZ error and echo cross-resonance gate (ECR) error (for the Eagle processor described below) represent the average gate fidelity over all possible inputs. These metrics exhibited no significant fluctuations throughout the course of our demonstrations.

To compare empirical results across different quantum processors, we also conducted demonstrations on IBM's previous-generation quantum processor called ``Eagle'' (ibm\_sherbrooke) \cite{Ibm_undated-ip,Ibm_undated-xp}. Unlike the Heron processor, the Eagle processor employs the ECR gate \cite{Sundaresan2020-yo} as a native two-qubit gate, although it has the same qubit connectivity as the Heron processor. The typical metrics of ibm\_sherbrooke over the demonstration period are shown in Table~\ref{table:processor_specifications} as well. Benchmarks of the Heron and Eagle processors are provided in Ref.~\cite{McKay2023-tj}.

\section{Demonstrations on the equality condition of the general QTUR}
\subsection{Measurement of the optimal observable $\mathcal{F}_{\rm opt}$}\label{section:appendix:Fopt-system}

\begin{figure*}[ht]
    \centering
    \subfloat[]{\includegraphics[width=0.4\textwidth]{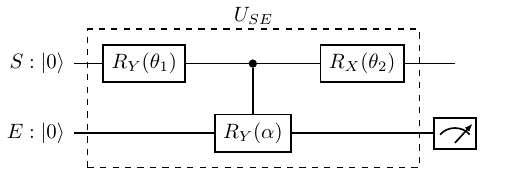}}\hspace{0\textwidth}
    \subfloat[]{\includegraphics[width=0.25\textwidth]{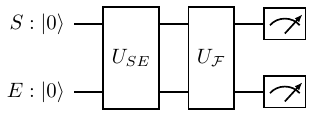}}\\
    \subfloat[]{\includegraphics[width=0.4\textwidth]{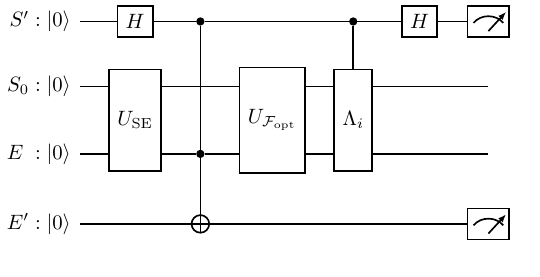}}\hspace{0.05\textwidth}
    \subfloat[]{\includegraphics[width=0.5\textwidth]{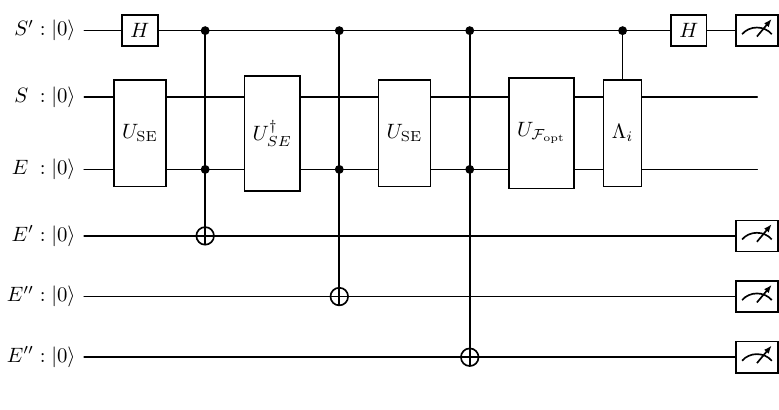}}
    \caption[Circuits for $\mathcal{F}_{\rm opt}$, $U_{SE}$, and $\mathcal{C}_{\mathcal{F}_{\rm opt}}$]{\label{fig:appendix:circuits-for-equality}Circuits for the optimal observable $\mathcal{F}_{\rm opt}$, the unitary operation $U_{SE}$ for a CPTP map, and the inherent-dynamics contribution $\mathcal{C}_{\mathcal{F}_{\rm opt}}$. (a) Parameterized circuit for implementing $U_{SE}$. (b) Circuit for measuring $\mathcal{F}_{\rm opt}$. (c) Optimized circuit for $C_1$. The controlled-CNOT gate, with control qubits denoted by black dots, is implemented using an approximate Toffoli gate. (d) Optimized circuit for $C_2$. Similar to (c), the controlled-CNOT gates are realized using approximate Toffoli gates.}
\end{figure*}

As described in the derivation of the general QTUR, equality is achieved for the observable $\mathcal{F}_{\rm opt}=\mathcal{L}=\ketbra{\Psi_{SE}(T)}-1/2(\ketbra*{\widetilde{\Psi}_{SE}(0)}{\Psi_{SE}(T)}+\ketbra*{\Psi_{SE}(T)}{\widetilde{\Psi}_{SE}(0)})$ when the initial state $\ket{\Psi_S(0)}$ is pure. In this demonstration, we investigate a method to measure the statistics of this observable, given a simple unitary operator $U_{SE}$. Specifically, we consider a two-qubit system, with one qubit representing the principal system $S$ and the other representing the environment $E$, both of which are initialized in the $\ket{0}$ state. The unitary operator $U_{SE}$ is defined as illustrated in Fig.~\ref{fig:appendix:circuits-for-equality}a. $U_{SE}$ comprises two single-qubit rotations around the Y-axis (RY) and X-axis (RX). Each rotation gate is parameterized by random variables $\theta_1$ and $\theta_2$ sampled from $[0,\pi/2]$. Additionally, $U_{SE}$ includes a controlled RY gate with the control qubit on $S$ and the target qubit on $E$. To modulate the interaction strength between $S$ and $E$, we use a parameter $\gamma$ chosen from the interval $[0,0.5]$, setting the rotation angle of the controlled RY gate to $\alpha=\pi\gamma$. With these parameters, we can theoretically obtain $\mathcal{L}$ and numerically diagonalize it as $\mathcal{L}=U_{\mathcal{F}_{\rm opt}}^\dagger\Lambda U_{\mathcal{F}_{\rm opt}}$, where $U_{\mathcal{F}_{\rm opt}}$ is the unitary matrix and $\Lambda$ is the diagonal matrix. This form of $\mathcal{L}$ can be interpreted as a Heisenberg picture of $\Lambda$ under unitary dynamics with $U_{\mathcal{F}_{\rm opt}}$. Consequently, we can access the statistics of $\mathcal{L}$, particularly its expectation and variance, by measuring $\Lambda$ with separated measurements on $S$ and $E$, as depicted in Fig.~\ref{fig:appendix:circuits-for-equality}b. Thus, the survival activity can be measured using the circuit in Fig.~\ref{fig:appendix:circuits-for-equality}a, while $\expval{\mathcal{L}}$ and ${\rm Var}[\mathcal{L}]$ can be obtained using the circuit in Fig.~\ref{fig:appendix:circuits-for-equality}b.

\subsection{Circuits for $\mathcal{C}_{\mathcal{F}_{\rm opt}}$\label{section:appendix:circuits-for-C-F-opt}}
Measurement of the inherent-dynamics contribution $\mathcal{C}_\mathcal{F}$ is based on Hadamard tests, as described in Appendix~\ref{section:appendix:C-F}. One notable difference between the circuits for $\mathcal{C}_{\mathcal{F}_{\rm opt}}$ and general $\mathcal{C}_\mathcal{F}$ is the inclusion of the operation $U_{\mathcal{F}_{\rm opt}}$ followed by controlled $\Lambda_i$ in the former. Here, $\Lambda_i$'s are Hermitian and unitary operators satisfying $\Lambda=\sum_i\Lambda_i$. Consequently, to obtain the value for $\Lambda$, we must execute circuits corresponding to each $\Lambda_i$ and sum the resulting values. Interestingly, we observe that for all the investigated circuits, $\Lambda$ can be empirically expressed in the form $\Lambda=c(\ketbra{00}-\ketbra{11})$, where $c$ is a scalar dependent on each $U_{SE}$. This peculiarity may be attributed to the structure of the $U_{SE}$ circuit depicted in Fig.~\ref{fig:appendix:circuits-for-equality}a. As a result of this property, we only need to implement circuits for $\Lambda_1=\sigma_z\otimes I_E$ and $\Lambda_2=I_S\otimes \sigma_z$.

The measurement of $\mathcal{C_1}$ and $C_2$ presents a challenge due to deep circuits. Therefore, as described in the main text, we optimize circuits by focusing on their algebraic structure and using approximate Toffoli gates. The approximate Toffoli gate requires only three CNOT gates, unlike the six CNOT gates needed for the exact Toffoli gate (note that these numbers refer to logical circuits). Remarkably, the approximate Toffoli gate behaves \textit{exactly} when the target is in $\ket{0}$, which is the case in our setup. We present the optimized circuit for $C_2$ in Fig.~\ref{fig:appendix:circuits-for-equality}d. 

\subsection{Depth evaluation\label{section:appendix:depth-evaluation-F-opt}}
\begin{figure}[ht]
    \centering
    \includegraphics[width=0.4\textwidth]{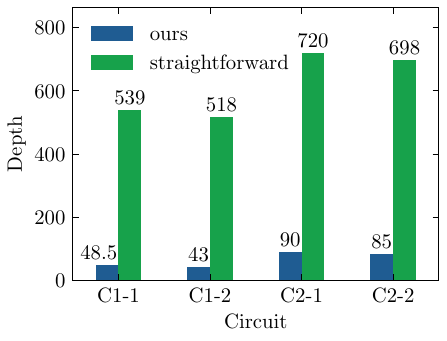}
    \caption[Depths of transpiled circuits for $\mathcal{C}_{\mathcal{F}_{\rm opt}}$]{\label{fig:appendix:depth-SLD}Depths of transpiled circuits for $C_1$ and $C_2$ regarding $\mathcal{F}_{\rm opt}$. The median depths of 50 circuits, corresponding to 50 different sets of parameters, are presented. The notation $C_{j{\rm -}i}$ denotes the circuit for $C_j$ associated with $\Lambda_i$.}
\end{figure}
The circuits designed thus far are logical circuits that ignore the physical constraints of qubits and operations. To execute these circuits on the quantum processor, they must be \textit{transpiled} into physical circuits that satisfy these constraints. In this study, we use \textit{Qiskit}, a standard Python library \cite{Qiskit_contributors_undated-lj}. Specifically, we employ the \textit{transpile} function with an \textit{optimization\_level} of 3 (i.e., the highest level). This function performs stochastic optimization to reduce the depth of the circuits, although it is not powerful enough to generate our optimized (logical) circuit in Fig.~\ref{fig:appendix:circuits-for-equality}c and \ref{fig:appendix:circuits-for-equality}d from the straightforward implementation in Figs.~\ref{fig:main:circuit-for-C1}b and \ref{fig:appendix:circuits-for-C2-naive}.

Figure~\ref{fig:appendix:depth-SLD} shows the depths of the transpiled circuits for $C_1$ and $C_2$ regarding 50 sets of parameters described in Appendix~\ref{section:appendix:Fopt-system}. While slight differences exist between the circuits for $\Lambda_1$ and $\Lambda_2$, our approach markedly reduces the overall depth. 

\subsection{Empirical results}\label{section:appendix:Fopt-empirical-results}
\begin{figure*}[ht]
    \centering
    \includegraphics[width=\textwidth]{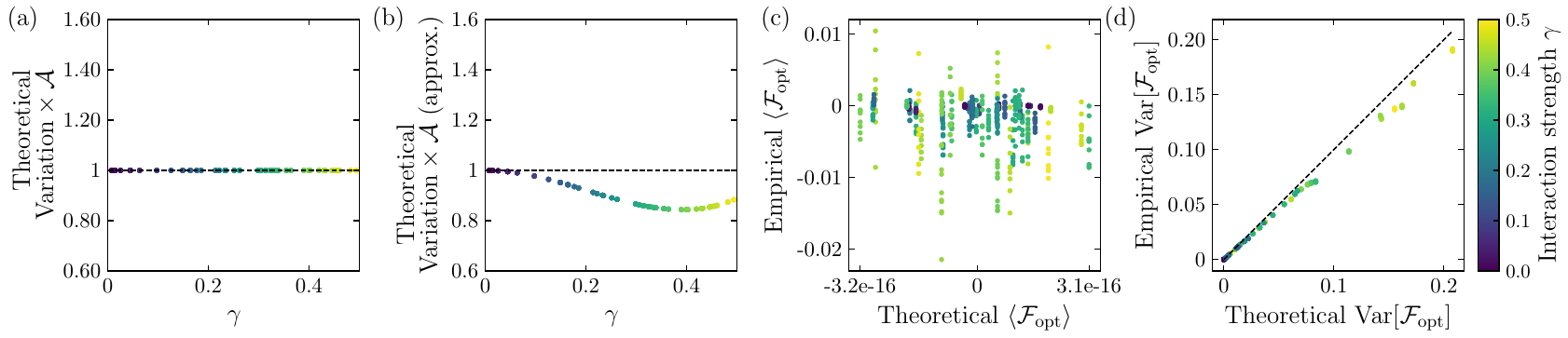}
    \caption[Theoretical, approximated, and empirical values regarding $\mathcal{F}_{\rm opt}$]{\label{fig:appendix:experiment-result-SLD-Heron-var}Comparison of the theoretical, approximated, and empirically obtained values regarding $\mathcal{F}_{\rm opt}$. The interaction strength $\gamma$ is represented by different colors. (a) Theoretical values of the product of the variation ${\rm Var}[\mathcal{F}_{\rm opt}]/(\expval{\mathcal{F}_{\rm opt}}-\mathcal{C}_{\mathcal{F}_{\rm opt}})^2$ and survival activity $\mathcal{A}$, plotted against $\gamma$. (b) Theoretically approximated values of the product of variation and $\mathcal{A}$. (c) Comparison of the theoretical and empirical values of the expectation value $\expval{\mathcal{F}_{\rm opt}}$. (d) Comparison of the theoretical and empirical values of the variance $\rm Var[\mathcal{F}_{\rm opt}]$. In (c) and (d), $50\times 10$ data points are plotted.}
\end{figure*}
\begin{figure*}[ht]
    \centering
    \includegraphics[width=\textwidth]{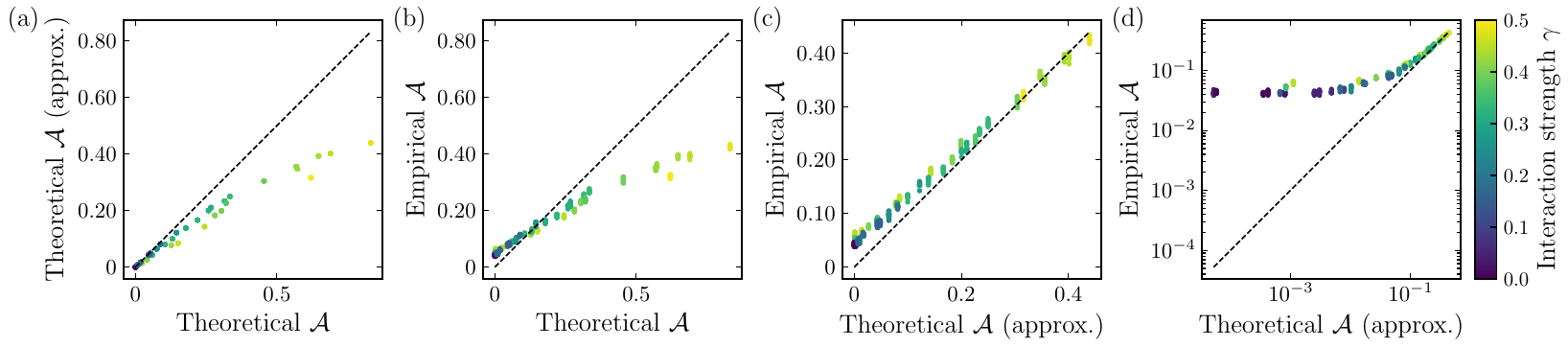}
    \caption[Theoretical, approximated, and empirical values of $\mathcal{A}$]{\label{fig:appendix:experiment-result-SLD-Heron-A}Comparison of theoretical, approximated, and empirical values of $\mathcal{A}$. (a) Theoretically approximated values plotted against theoretical values. (b) Empirical values plotted against theoretical values. (c) Empirical values plotted against theoretically approximated values. (d) Logarithmic plot of the data presented in (c). For panels (b)--(d), $50\times 10$ data points are displayed.}
\end{figure*}
\begin{figure*}[ht]
    \centering
    \includegraphics[width=\textwidth]{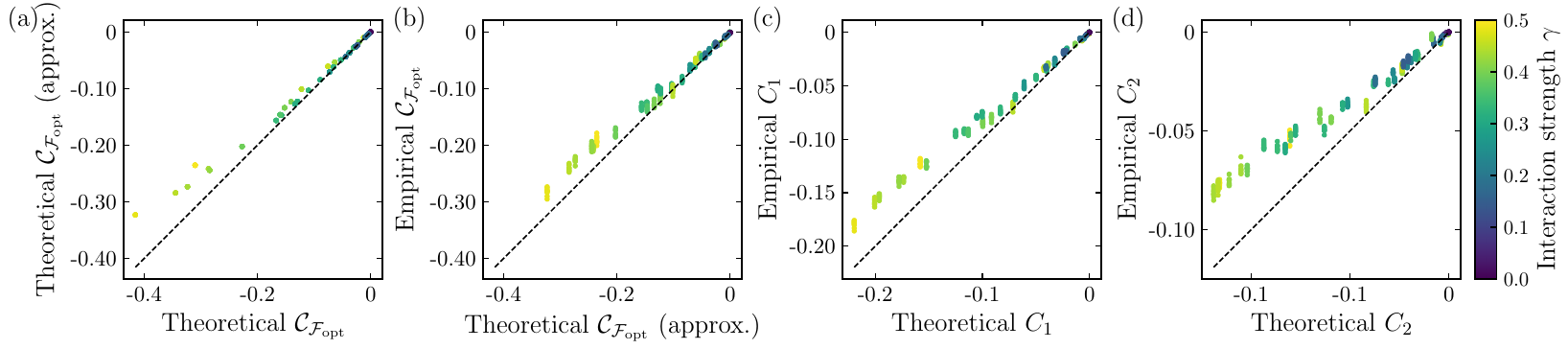}
    \caption[Theoretical, approximated, and empirical values of $\mathcal{C}_{\mathcal{F}_{\rm opt}}$]{\label{fig:appendix:experiment-result-SLD-Heron-CF}Comparison of theoretical, approximated, and empirical values of $\mathcal{C}_{\mathcal{F}_{\rm opt}}$. (a) Theoretically approximated values of $\mathcal{C}_{\mathcal{F}_{\rm opt}}$ plotted against theoretical values. (b) Empirical values of $\mathcal{C}_{\mathcal{F}_{\rm opt}}$ plotted against theoretically approximated values. (c) Empirical values of $C_1$ plotted against theoretical values. (d) Empirical values of $C_2$ plotted against theoretical values. For panels (b)--(d), $50\times 10$ data points are displayed.}
\end{figure*}
We conducted demonstrations on 50 sets of parameter combinations for $\theta_i$ and $\gamma$. For each combination, we performed 8,000 measurements on the corresponding circuits. This procedure was repeated 10 times for the Heron processor and 5 times for the Eagle processor, which is described later.

Figure~\ref{fig:appendix:experiment-result-SLD-Heron-var} shows the influence of the theoretical approximation of $\mathcal{A}$ and $\mathcal{C}_\mathcal{F}$, as well as the empirical values of $\expval{\mathcal{F}_{\rm opt}}$ and ${\rm Var}[\mathcal{F}_{\rm opt}]$. As shown in Fig.~\ref{fig:appendix:experiment-result-SLD-Heron-var}a, our design of $\mathcal{F}_{\rm opt}$ achieves equality in the general QTUR. The influence of the theoretical approximation is depicted in Fig.~\ref{fig:appendix:experiment-result-SLD-Heron-var}b, where the product of variation and $\mathcal{A}$ falls below one, particularly for relatively large $\gamma$ values. Nevertheless, we conclude that the deviation to be insignificant compared to the fluctuation of the empirical values, as demonstrated in Fig.~\ref{fig:experiment-result-Fopt}. Furthermore, Fig.~\ref{fig:appendix:experiment-result-SLD-Heron-var}c reveals that the empirical values of $\expval{\mathcal{F}_{\rm opt}}$ are nearly zero, aligning with the theoretical prediction of $\expval{\mathcal{F}_{\rm opt}}=0$ due to the property of the symmetric logarithmic derivatives. The variance of $\mathcal{F}_{\rm opt}$ is presented in Fig.~\ref{fig:appendix:experiment-result-SLD-Heron-var}d, where the empirical values exhibit strong agreement with the predictions.

Figure~\ref{fig:appendix:experiment-result-SLD-Heron-A} presents the results of the survival activity $\mathcal{A}$. As seen in Fig.~\ref{fig:appendix:experiment-result-SLD-Heron-A}a, the approximated values of $\mathcal{A}$ align with the exact values for small $\gamma$ but are smaller than the exact values for relatively large $\gamma$. This deviation accounts for the deviation between the theoretical predictions and the limit in Fig.~\ref{fig:appendix:experiment-result-SLD-Heron-var}b. Nevertheless, we emphasize again that the derivation in Fig.~\ref{fig:appendix:experiment-result-SLD-Heron-var}b, arising from the approximation, is \textit{smaller than the fluctuations of the empirical data} around the theoretical predictions. Then, Fig.~\ref{fig:appendix:experiment-result-SLD-Heron-A}b reveals that the empirical values of $\mathcal{A}$ follow the same trend as the approximated theoretical values in Fig.~\ref{fig:appendix:experiment-result-SLD-Heron-A}a. However, we observe that the empirical values slightly surpass the predictions for small $\gamma$, which is more clearly illustrated in Figs.~\ref{fig:appendix:experiment-result-SLD-Heron-A}c and \ref{fig:appendix:experiment-result-SLD-Heron-A}d. We attribute the deviation of the empirical values of the product of variation and survival activity from the limit in Fig.~\ref{fig:experiment-result-Fopt}a to this discrepancy. We analyze the cause of this empirical error in the next subsection.

Figure~\ref{fig:appendix:experiment-result-SLD-Heron-CF} displays the results of the inherent-dynamics contribution $\mathcal{C}_{\mathcal{F}_{\rm opt}}$. As shown in Fig.~\ref{fig:appendix:experiment-result-SLD-Heron-CF}a, the approximated values of $\mathcal{C}_{\mathcal{F}_{\rm opt}}$ are slightly exceed the theoretical values for large $\gamma$. This trend is mirrored by the empirical values in Figs.~\ref{fig:appendix:experiment-result-SLD-Heron-CF}b, \ref{fig:appendix:experiment-result-SLD-Heron-CF}c, and \ref{fig:appendix:experiment-result-SLD-Heron-CF}d.

\subsection{Analysis of the empirical errors}\label{section:appendix:analyze-errors-Fopt}
As a major cause of the discrepancies between empirical and theoretical values of $\mathcal{A}$, we consider the states of the environments. In particular, we focus on the temperature of environments, which should be finite in practice. To assess the thermal configuration of qubits, we examine the calibration data of the target processor. The data offers the empirical values of ``readout errors'' as described in Appendix~\ref{section:appendix:IBM}. This value includes both state preparation errors (population of excited state $\ket{1}$ in qubits intended to be $\ket{0}$) and true readout errors (measurement errors of qubits in known states). While these state preparation and measurement (SPAM) errors are generally inseparable \cite{Jayakumar2024-la}, we follow previous studies \cite{Solfanelli2021-rc,Buffoni2022-jf,Bassman-Oftelie2024-nf} and interpret the ``readout error'' as the effective population of the excited state in equilibrated qubits. By incorporating these thermal states, we revise our theoretical predictions for $\mathcal{F}_{\rm opt}$, as shown in Figs.~\ref{fig:experiment-result-Fopt}a and \ref{fig:experiment-result-Fopt}c in the main text.

\begin{figure*}[ht]
    \centering
    \includegraphics[width=0.8\textwidth]{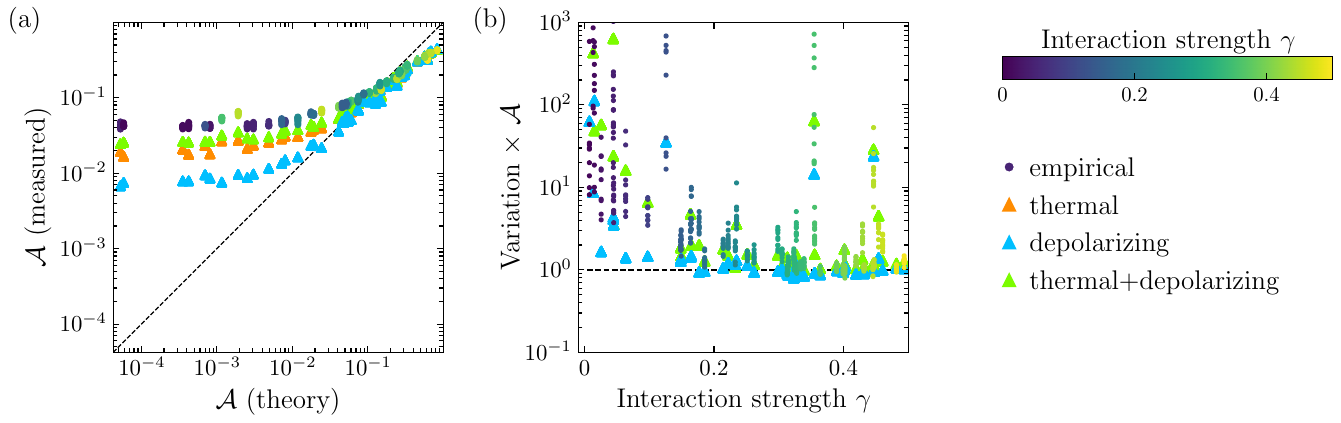}
    \caption[Noisy simulations for $\mathcal{F}_{\rm opt}$]{\label{fig:appendix:experiment-result-SLD-Heron-thermal}Comparison between empirical results and noisy simulations for $\mathcal{F}_{\rm opt}$. (a) Empirical and simulated values of $\mathcal{A}$ versus theoretical predictions. (b) Product of the variation ${\rm Var}[\mathcal{F}_{\rm opt}]/(\expval{\mathcal{F}_{\rm opt}}-\mathcal{C}_{\mathcal{F}_{\rm opt}})^2$ and $\mathcal{A}$ versus interaction strength $\gamma$.}
\end{figure*}

To substantiate the dominant role of thermal effects, we compare them with gate errors through numerical simulations. Using Qiskit \cite{Qiskit_contributors_undated-lj}, we simulate the transpiled physical circuits incorporating both thermal initial states and depolarizing errors. While gate parameters are set according to the calibration data, we exclude RZ gate errors due to unavailable error rates. The simulation results, presented in Fig.~\ref{fig:appendix:experiment-result-SLD-Heron-thermal}, show that depolarizing errors alone cannot explain the discrepancies between theory and empirical data, highlighting the necessity of considering thermal initial states.

Next, we estimate the effective temperature of qubits from the effective equilibrium population. The Gibbs state of qubits at temperature $\mathcal{T}$ is given by
\begin{align}
    \rho_{\rm Gibbs}=\frac{1}{Z}e^{-\beta (E_0\ketbra{0}+E_1\ketbra{1})},
\end{align}
with the inverse temperature $\beta=1/k_{\rm B}\mathcal{T}$. To determine $\Delta E=E_1-E_0$, we refer to the qubit resonance frequencies reported by IBM \cite{Ibm_undated-ip}. Although these frequencies are not published for the Heron processor, we use data from the previous Eagle processor as a reasonable approximation, which shows frequencies in the range of 4.5--5 GHz. Based on the excited population, we estimate an effective temperature of approximately 60~mK, consistent with values reported in Refs.~\cite{Buffoni2022-jf,Bassman-Oftelie2024-nf}. The precise effective temperature is not critical for our analysis and this estimate helps interpret our demonstration results in terms of thermal effects.

\subsection{Comparison to the straightforward implementation for $\mathcal{C}_\mathcal{F}$}\label{section:appendix:results-for-straightforward}
\begin{figure*}[ht]
    \centering
    \includegraphics[width=\textwidth]{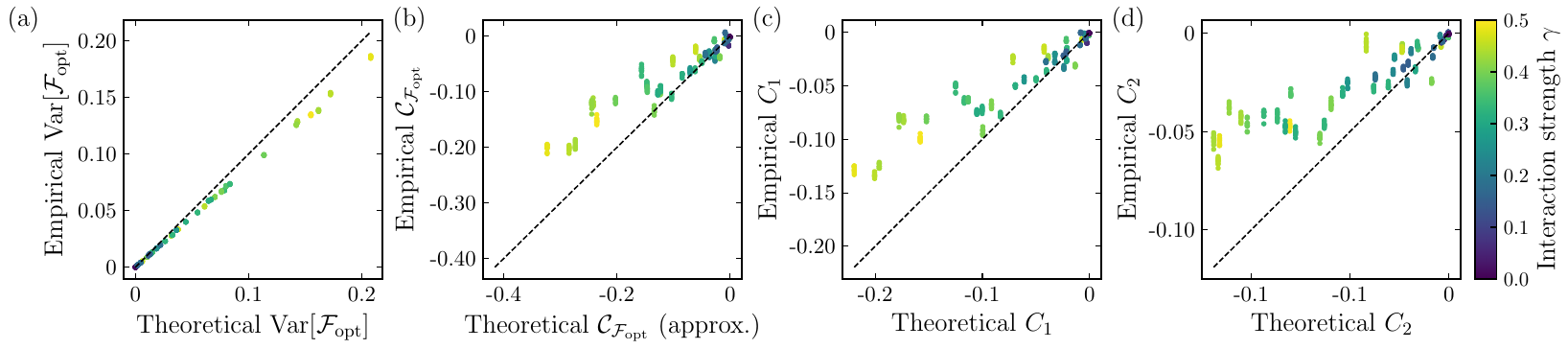}
    \caption[Results for $\mathcal{F}_{\rm opt}$ with the straightforward implementation]{\label{fig:appendix:experiment-result-SLD-Heron-naive-CF}Comparison of theoretical, approximated, and empirical results regarding $\mathcal{F}_{\rm opt}$ using the straightforward implementation. (a) Empirical values of ${\rm Var}[\mathcal{F}_{\rm opt}]$ plotted against theoretical values. (b) Empirical values of $\mathcal{C}_{\mathcal{F}_{\rm opt}}$ plotted against theoretically approximated values. (c) Empirical values of $C_1$ plotted against theoretical values. (d) Empirical values of $C_2$ plotted against theoretical values. For panels (b)--(d), $50\times 10$ data points are displayed.}
\end{figure*}
In Fig.~\ref{fig:appendix:experiment-result-SLD-Heron-naive-CF}, we present the results of the straightforward implementation of the Hadamard test for $\mathcal{C}_\mathcal{F}$. Fig.~\ref{fig:appendix:experiment-result-SLD-Heron-naive-CF}a, which displays ${\rm Var}[\mathcal{F}_{\rm opt}]$, confirms that the qubit and gate quality for this demonstration is comparable to that of the previous demonstrations. However, Fig.~\ref{fig:appendix:experiment-result-SLD-Heron-naive-CF}b reveals that the empirical values of $\mathcal{C}_{\mathcal{F}_{\rm opt}}$ are noisy and scattered compared to the predicted values. This discrepancy can be attributed to the fact that the straightforward implementation requires circuits with approximately ten times the depth of the optimized circuits, leading to increased noise susceptibility. Figures~\ref{fig:appendix:experiment-result-SLD-Heron-naive-CF}c and \ref{fig:appendix:experiment-result-SLD-Heron-naive-CF}d illustrate that $C_2$, which requires longer circuits than $C_1$, is more severely affected by noise. These findings stand in sharp contrast to the results obtained using the optimized circuits, as shown in Fig.~\ref{fig:appendix:experiment-result-SLD-Heron-CF}.

\subsection{Comparison to the Eagle processor}\label{section:appendix:results-for-Eagle}
\begin{figure*}[ht]
    \centering
    \includegraphics[width=\textwidth]{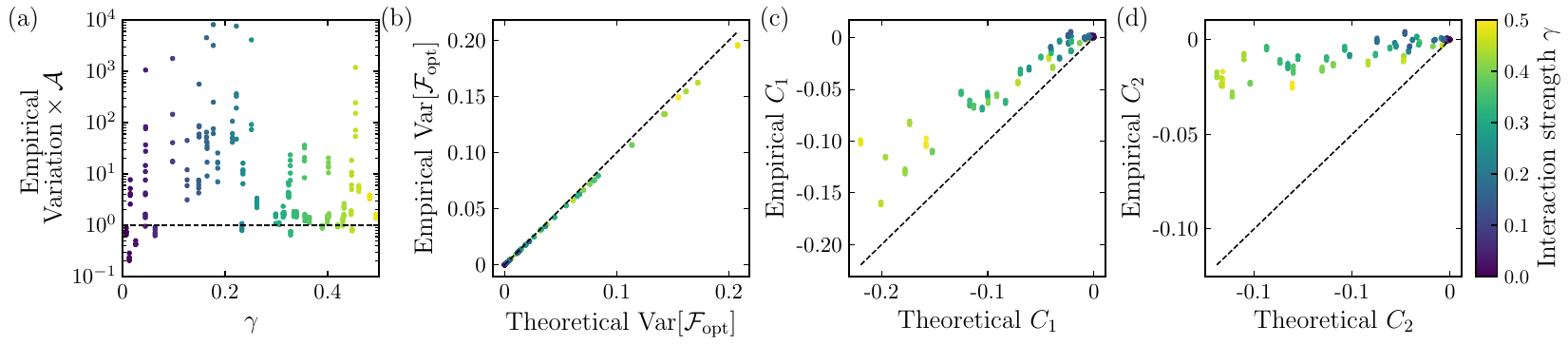}
    \caption[Results for $\mathcal{F}_{\rm opt}$ on the Eagle processor]{\label{fig:appendix:experiment-result-SLD-Eagle-var}Results on $\mathcal{F}_{\rm opt}$ obtained with the Eagle processor. (a) Empirical values of the product of the variance and $\mathcal{A}$ plotted against the interaction strength $\gamma$. (b) Empirical values of ${\rm Var}[\mathcal{F}_{\rm opt}]$ plotted against theoretical values. (c) Empirical values of $C_1$ plotted against theoretical values. (d) Empirical values of $C_2$ plotted against theoretical values. For all panels, $50\times 5$ data points are displayed.}
\end{figure*}
To compare the empirical results on the equality condition of the general QTUR across different quantum processors, we conducted the same demonstrations using the Eagle processor. Figure~\ref{fig:appendix:experiment-result-SLD-Eagle-var} presents the results of these demonstrations. As shown in Fig.~\ref{fig:appendix:experiment-result-SLD-Eagle-var}a, the empirical values violate the general QTUR. Figure~\ref{fig:appendix:experiment-result-SLD-Eagle-var}b reveals that the empirical values of ${\rm Var}[F_{\rm opt}]$ align well with the predictions, and the difference between the processors is not apparent. However, Figs.~\ref{fig:appendix:experiment-result-SLD-Eagle-var}c and \ref{fig:appendix:experiment-result-SLD-Eagle-var}d demonstrate that the empirical values of $C_1$ and $C_2$ on the Eagle processor are significantly noisier and deviate from the predictions. These findings suggest that the difference in processor accuracy plays a crucial role in demonstrations involving the general QTUR, particularly when investigating its equality condition.

\section{Demonstrations on the quantum time correlator $R(T)$}
\subsection{Target system $S$ and CPTP map $\Phi$\label{section:appendix:corr-system}}
\begin{figure*}[ht]
    \centering
    \includegraphics[width=0.8\textwidth]{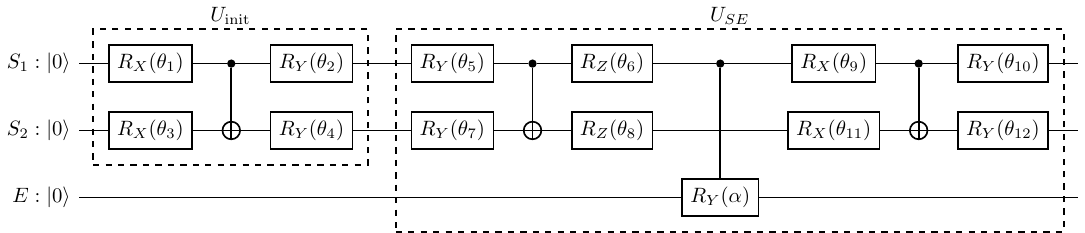}
    \caption[Circuits for state preparation and a CPTP map for time correlators]{\label{fig:state-CPTP-circuit}Circuits for preparing the initial state $\rho_{S}(0)$ and constructing the CPTP map $\Phi$. The initial state $\rho_{S}(0)$ is generated by applying a randomly parameterized unitary gate $U_{\rm init}$ to the ground state $\ket{0}\otimes\ket{0}$ of the system $S$. The CPTP map $\Phi$ acting on the system $S$ is realized by applying a unitary operator $U_{SE}$ to the composite system $S$+$E$, where $U_{SE}$ comprises randomly parameterized rotation gates.
    }
\end{figure*}
In this demonstration, we consider a principal system $S$ comprising two qubits and an environment with one qubit. Figure~\ref{fig:state-CPTP-circuit} illustrates the unitary operator $U_{\rm init}$ for state preparation and $U_{SE}$ for a CPTP map $\Phi$ on $S$. The operator $U_{\rm init}$ comprises two RX gates, one CNOT gate, and two RY gates. Each rotation gate is parameterized by $\theta_i$ for $i=1,\ldots,4$, which is randomly selected from the interval $[0,\pi/2]$. Similarly, $U_{SE}$ comprises single-qubit rotation gates randomly parameterized by $\theta_i$ for $i=5,\ldots,12$, chosen from the same interval. Additionally, $U_{SE}$ includes a controlled RY gate with the control qubit on $S$ and the target qubit on $E$. To modulate the interaction strength between $S$ and $E$, we use a parameter $\gamma$ chosen from the interval $[0,0.5]$, setting the rotation by the controlled RY gate to $\alpha=\pi\gamma$. In the following demonstrations, we fix and investigate 50 sets of the parameters $\{\theta_i\}$ and $\gamma$.

\subsection{Details of the circuit for measuring $R(T)$ illustrated in Fig.~\ref{fig:illustration}b}\label{section:appendix:details-circuits-for-Fig1b}
\begin{figure}[ht]
    \centering
    \includegraphics[width=0.45\textwidth]{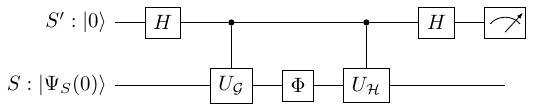}
    \caption[Protocol for measuring quantum time correlators]{\label{fig:appendix:timecorrelator}Schematic representation of the protocol for measuring quantum time correlators. The ancillary system \textit{S'} is a qubit, while the principal system \textit{S} undergoes a CPTP map $\Phi$. The map is interspersed between the controlled unitary operations $U_\mathcal{G}^{(c)}$ and $U_\mathcal{H}^{(c)}$, which are applied to the combined system \textit{S'} and \textit{S}.}
\end{figure}

We focus on the two-point time correlator $R(t) = \langle \mathcal{H}(t)\mathcal{G}(0) \rangle_{\rho_{S}(0)}$ of unitary observables $\mathcal{G}$ and $\mathcal{H}$, where $\mathcal{H}(t)=\sum_{m} V_m^{\dagger} \mathcal{H}V_m$ is the Heisenberg picture of $\mathcal{H}$ at time $t$. We apply the general QTUR to a circuit for the protocol to measure $R(t)$, which is illustrated in Fig.~\ref{fig:illustration}b. This protocol, based on Ref.~\cite{Pedernales2014-ui}, employs the Hadamard test described above. The main part of the circuit for the protocol is shown in Fig.~\ref{fig:appendix:timecorrelator} and can be described as follows:
\begin{enumerate}
    \item Initialize an auxiliary qubit $S'$ in the state $\ket{0}$.
    \item Apply an Hadamard gate to $S'$ to obtain the state $\displaystyle\ket{+}=\frac{1}{\sqrt{2}}(\ket{0}+\ket{1})$.
    \item Perform a controlled unitary operation $U_\mathcal{G}^{(c)}$ on the combined system $S+S'$, which applies $\mathcal{G}$ on $S$ conditioned on the state of $S'$.
    \item Subject the main system $S$ to the CPTP map $\Phi$.
    \item Perform a controlled unitary operation $U_\mathcal{H}^{(c)}$ on $S+S'$, which applies $\mathcal{H}$ on $S$ conditioned on the state of $S'$.
    \item Measure the ancilla $S'$ to obtain the expectation values of the Pauli operators $\sigma_x$ and $\sigma_y$, which correspond to the real and imaginary components of $R(t)$, respectively. Note that due to hardware constraints, the $\sigma_x$ measurement is performed by applying an Hadamard gate followed by a $\sigma_z$ measurement.
\end{enumerate}
Let the final state of the composite system prior to measurement be denoted as $\ket{\Psi_{S'S}(T)}$. The time correlator $R(t)$ satisfies
\begin{align}
    R(T) &= {\rm Tr}[\rho_{S}(0)\mathcal{H}(T)\mathcal{G}(0)]\nonumber\\
    &=\ev{\sigma_x\otimes I_{S}}{\Psi_{S'S}(T)}\nonumber\\
    &\quad+i \ev{\sigma_y\otimes I_{S}}{\Psi_{S'S}(T)}.
\end{align}
Therefore, by measuring the Pauli operators $\sigma_x$ and $\sigma_y$ on the ancillary system $S'$, we can extract the real and imaginary components of $R(T)$, respectively. In the following, we focus on the real component ${\rm Re}[R(T)]$.

\subsection{Application of the general QTUR in Eq.~\eqref{eq:GeneralTUR}}\label{section:appendix:derivation-of-RT-precision-bound}
We apply the general QTUR [Eq.~\eqref{eq:GeneralTUR}] to investigate the precision and involved thermodynamic quantity of the protocol described above. The generality of the QTUR allows us to consider arbitrary observables. As depicted in Fig.~\ref{fig:illustration}b, we treat the intermediate state $\rho_{S'S}^\mathcal{G}=U_\mathcal{G}^{(c)} (\ketbra{+}\otimes\rho_{S}(0)) {U_\mathcal{G}^{(c)\dagger}}$ as the initial state and the observable $\mathcal{F}_{\rm corr}=U_\mathcal{H}^{(c)}(\sigma_x\otimes I_S)U_\mathcal{H}^{(c)\dagger}$ as the observable to be bounded, which satisfies $\expval{\mathcal{F}_{\rm corr}}={\rm Re}[R(T)]$. The survival activity for the CPTP map $\Phi$ is given by $\mathcal{A}=\operatorname{Tr}[\rho_{S'S}^\mathcal{G}(V_0^\dagger V_0)^{-1}]-1$, and the inherent-dynamics contribution is
\begin{align}
    \mathcal{C}_{\mathcal{F}_{\rm corr}}&={\rm Re}[\mel*{\widetilde{\Psi}_{SE}(0)}{\mathcal{F}_{\rm corr}}{\Psi_{SE}(T)}]\nonumber\\
    &={\rm Re}[{\rm Tr}[\rho_{S'S}^{V_0}(V_0V_0^\dagger)^{-1}\mathcal{F}_{\rm corr}]],
\end{align}    
where $\rho_{S'S}^{V_0}=(I_{S'}\otimes V_0)\rho_{S'S}^\mathcal{G} (I_{S'}\otimes V_0^\dagger)$. We note that the derivation of $\mathcal{C}_{\mathcal{F}_{\rm corr}}$ is based on $\mathcal{C}_{\mathcal{F}}$ for $\mathcal{F}_{S}$ in Eq.~\eqref{eq:appendix:C-Fs}. For these quantities, the general QTUR holds as
\begin{align}
    \frac{{\rm Var}[\mathcal{F}_{\rm corr}]}{\qty(\expval{\mathcal{F}_{\rm corr}}-\mathcal{C}_{\mathcal{F}_{\rm corr}})^2}\geq\frac{1}{\mathcal{A}}.
\end{align}
Using ${\rm Re}[R(T)]=\expval{\mathcal{{F}_{\rm corr}}}$ and defining ${\rm Var}[{\rm Re}[R(T)]]={\rm Var}[\mathcal{F}_{\rm corr}]$, we obtain
\begin{align}
    \frac{{\rm Var}[{\rm Re}[R(T)]]}{\qty({\rm Re}[R(T)]-\mathcal{C}_{\mathcal{F}_{\rm corr}})^2}\geq\frac{1}{\mathcal{A}}.\label{eq:appendix:generalTUR-corr}
\end{align}

\subsection{Derivation of the trade-off relation on quantum time correlators}\label{section:appendix:derivation-of-RT-tradeoff}
Starting from the general QTUR on $R(t)$ in Eq.~\eqref{eq:appendix:generalTUR-corr}, we derive a trade-off relation beyond precision. Clearing the denominator of Eq.~\eqref{eq:appendix:generalTUR-corr} yields
\begin{align}
    |{\rm Re}[R(T)]-\mathcal{C}_{\mathcal{F}_{\rm corr}}|\leq\sqrt{{\rm Var}[{\mathcal{F}_{\rm corr}}]\mathcal{A}}.\label{eq:appendix:correlationbound-first}
\end{align}
We proceed to evaluate the quantities in Eq.~\eqref{eq:appendix:correlationbound-first}. The observable ${\mathcal{F}_{\rm corr}}$ is defined as $\mathcal{F}_{\rm corr}=U_\mathcal{H}^{(c)}(\sigma_x\otimes I_S)U_\mathcal{H}^{(c)\dagger}$, where the controlled unitary operation $U_\mathcal{H}^{(c)}$ is given by $U_\mathcal{H}^{(c)}=\ketbra{0}\otimes I+\ketbra{1}\otimes \mathcal{H}$. Substituting this expression, we obtain ${\mathcal{F}_{\rm corr}}=\sigma_x\otimes \mathcal{H}$, with $\mathcal{H}$ being unitary and Hermitian by definition. To emphasize the dependence on $\mathcal{G}$ and $\mathcal{H}$, we introduce the inherent-dynamics contribution $\mathcal{C}_{\mathcal{G},\mathcal{H}}$, which is equivalent to $\mathcal{C}_{\mathcal{F}_{\rm corr}}$, as follows:

\begin{align}
    \mathcal{C}_{\mathcal{G},\mathcal{H}}={\rm Re}[{\rm Tr}[\rho_{S'S}^{V_0}(I_{S'}\otimes(V_0V_0^\dagger)^{-1})(\sigma_x\otimes \mathcal{H})]].
\end{align}
$\mathcal{C}_{\mathcal{G},\mathcal{H}}$ is further calculated as
\begin{widetext}
\begin{align}
    \mathcal{C}_{\mathcal{G},\mathcal{H}}
    &={\rm Re}[{\rm Tr}[\rho_{S'S}^{\mathcal{G}}(\sigma_x\otimes V_0^{-1} \mathcal{H}V_0)]]\nonumber\\
    &={\rm Re}[{\rm Tr}[\ketbra{+}\otimes\rho_S(0)(\ketbra{0}{1}\otimes V_0^{-1}\mathcal{H}V_0\mathcal{G}+\ketbra{1}{0}\otimes \mathcal{G} V_0^{-1}\mathcal{H}V_0)]]\nonumber\\
    &={\rm Re}[{\rm Tr}[\frac{1}{2}(\ketbra{0}{1}+\ketbra{1})\otimes \rho_S(0)V_0^{-1}\mathcal{H}V_0\mathcal{G}+\frac{1}{2}(\ketbra{0}+\ketbra{1}{0})\otimes\rho_S(0)\mathcal{G} V_0^{-1}\mathcal{H}V_0]]\nonumber\\
    &=\frac{1}{2}{\rm Re}[{\rm Tr}[\rho_S(0)(V_0^{-1}\mathcal{H}V_0\mathcal{G}+\mathcal{G}V_0^{-1}\mathcal{H}V_0)]]\nonumber\\
    &=\frac{1}{2}{\rm Re}[\expval{V_0^{-1}\mathcal{H}V_0\mathcal{G}+\mathcal{G}V_0^{-1}\mathcal{H}V_0}]\label{eq:appendix:C-GH}
\end{align}
\end{widetext}

Next, the upper bound for the variance of $\mathcal{F}_{\rm corr}$, i.e., the variance of ${\rm Re}[R(T)]$, is calculated as
\begin{align}
    {\rm Var}[\mathcal{F}_{\rm corr}]
    &=\expval{\mathcal{F}_{\rm corr}^2}-\expval{\mathcal{F}_{\rm corr}}^2\nonumber\\
    &\leq\expval{\mathcal{F}_{\rm corr}^2}
    =\expval{\mathcal{F}_{\rm corr}^\dagger\mathcal{F}_{\rm corr}}
    =1.\label{eq:appendix:corr-var-upper}
\end{align}
Finally, by substituting Eqs.~\eqref{eq:appendix:C-GH} and \eqref{eq:appendix:corr-var-upper} and $\expval{\mathcal{F}_{\rm corr}}={\rm Re}[R(T)]$ into Eq.~\eqref{eq:appendix:correlationbound-first}, we obtain Eq.~\eqref{eq:timeCorrelationBound}.

In the weak coupling limit where $V_0^\dagger V_0$ satisfies $V_0^\dagger V_0=I_S-\epsilon$, we performed a perturbative expansion of $\mathcal{C}_{\mathcal{G},\mathcal{H}}$ in the same way as for $\mathcal{C}_{\mathcal{F}_S}$. $\expval{V_0^{-1}\mathcal{H}V_0\mathcal{G}}$ reads
\begin{align}
    \expval{V_0^{-1}\mathcal{H}V_0\mathcal{G}}
    &\approx \expval{(I-\frac{1}{2}\epsilon+\epsilon)U_0^\dagger \mathcal{H}V_0\mathcal{G}}\nonumber\\
    &=\expval{V_0^\dagger \mathcal{H}V_0\mathcal{G}}+\expval{\epsilon U_0^\dagger \mathcal{H}V_0\mathcal{G}}\nonumber\\
    &=\expval{V_0^\dagger \mathcal{H}V_0\mathcal{G}}+\expval{\epsilon U_0^\dagger \mathcal{H}U_0\mathcal{G}}.
\end{align}
$\expval{\mathcal{G}V_0^{-1}\mathcal{H}V_0}$ is similarly calculated as
\begin{align}
    \expval{\mathcal{G}V_0^{-1}\mathcal{H}V_0}&\approx \expval{\mathcal{G}V_0^\dagger \mathcal{H}V_0}+\expval{\mathcal{G}\epsilon U_0^\dagger\mathcal{H}U_0}.
\end{align}
Therefore, $\mathcal{C}_{\mathcal{G},\mathcal{H}}$ is approximated as
\begin{align}
    \mathcal{C}_{\mathcal{G},\mathcal{H}}\approx{\rm Re}[\expval*{\mathcal{G}V_0^\dagger \mathcal{H}V_0}]+\frac{1}{2}{\rm Re}[\expval*{\epsilon U_0^\dagger \mathcal{H}U_0\mathcal{G}}+\expval*{\mathcal{G}\epsilon U_0^\dagger\mathcal{H}U_0}].
\end{align}
The primary term is a time correlator regarding $\mathcal{G}$ and $\mathcal{H}$ when the system $S$ is evolved by non-trace-preserving dynamics $V_0$.

\subsection{Circuits for $\mathcal{A}$ and $\mathcal{C}_{\mathcal{F}_{\rm corr}}$\label{section:appendix:circuits-for-Ag-and-C-F}}
We present the circuits used to measure the survival activity $\mathcal{A}$ and the inherent-dynamics contribution $\mathcal{C}_{\mathcal{F}_{\rm corr}}$. To estimate $\mathcal{A}$, we employ the approximation $\mathcal{A}\approx 1-p_0$, where $p_0={\rm Tr}[\rho_{S'S}^{\mathcal{G}}V_0^\dagger V_0]$, obtained by truncating Eq.~\eqref{eq:appendix:xi-expansion} at $N=1$. The probability $p_0$ is estimated using the same circuit utilized to measure ${\rm Re}[R(T)]$, as depicted in Fig.~\ref{fig:illustration}b, by measuring the environmental qubit $E$.

\begin{figure*}[ht]
    \centering
    \includegraphics[width=0.6\textwidth]{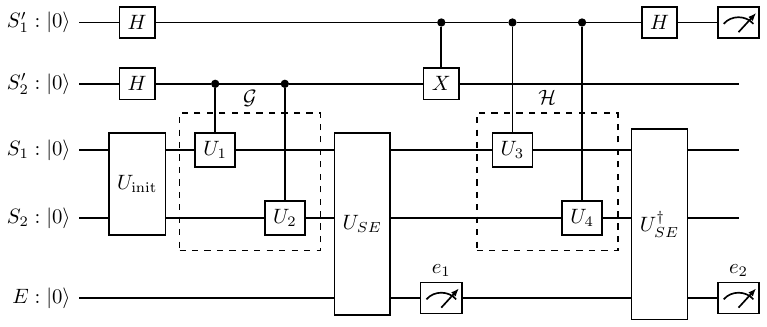}
    \caption[Optimized circuit for measuring $\mathcal{C}_{\mathcal{F}_{\rm corr}}$]{\label{fig:circuit-corr-rhoFV-optimized}Optimized circuit for measuring the $C_2$ term in $\mathcal{C}_{\mathcal{F}_{\rm corr}}$. The ancilla qubit $S'_2$ is used to create the time correlator as shown in Fig.~\ref{fig:illustration}b, while the ancilla qubit $S'_1$ is employed for the Hadamard test to measure the correlator in $C_2$. The circuit has the final measurement on $S_1'$, the mid-circuit measurement denoted by $e_1$, and the final measurement on the environment $E$ denoted by $e_2$. The unitary operations $U_i$ for $i=1,\dots,4$ are determined by the specific choice of $\mathcal{G}$ and $\mathcal{H}$ operators.}
\end{figure*}
Measuring $\mathcal{C}_{\mathcal{F}_{\rm corr}}\approx 2C_1-C_2$, where $C_1$ and $C_2$ are defined in Eqs.\eqref{eq:appendix:C-1-def} and \eqref{eq:appendix:C-2-def}, is the challenging part of this demonstration. As discussed in Appendix~\ref{section:appendix:C-F}, $\mathcal{C}_{\mathcal{F}_{\rm corr}}$ is determined through Hadamard tests for $C_1$ and $C_2$. However, the straightforward implementation of these tests requires long circuits, which can be problematic. To mitigate this issue, we leverage the simplified form of $\mathcal{C}_{\mathcal{F}}$ for the local observable $\mathcal{F}_{S}$, which includes $\mathcal{F}_{\rm corr}$, as shown in Eq.~\eqref{eq:appendix:C-Fs}. This simplification allows us to avoid the use of controlled unitary operations involving three or more qubits. The circuit for measuring $C_2$ is presented in Fig.~\ref{fig:circuit-corr-rhoFV-optimized}, which, unlike the straightforward implementation in Fig.~\ref{fig:appendix:circuits-for-C2-naive}, does not require a controlled $U_{SE}$ operation. Furthermore, we employ the mid-circuit measurement for the environment, denoted by $e_1$ in Fig.~\ref{fig:circuit-corr-rhoFV-optimized}, which reduces the number of qubits needed and, consequently, the number of resource-intensive swap operations. We remark that mid-circuit measurements take time to complete. However, in the circuit shown in Fig.~\ref{fig:circuit-corr-rhoFV-optimized}, this is not an issue because the measurement can be performed concurrently with other operations involving $S'_1$, $S'_2$, $S_1$, and $S_2$. This is not the case for the straightforward circuit in Fig.~\ref{fig:appendix:circuits-for-C2-naive}, where there is insufficient time for the mid-circuit measurement of the environment. A quantitative comparison of the depth of the straightforward and optimized circuits for $C_2$ is provided in the next subsection. Remarkably, the same circuit used for measuring ${\rm Re}[R(T)]$, shown in Fig.~\ref{fig:illustration}b, can also be used to measure $C_1$ due to the simplified form of $\mathcal{C}_{\mathcal{F}_{\rm corr}}$. In summary, we employ two circuits in this demonstration: one for measuring ${\rm Re}[R(T)]$, $\mathcal{A}$, and $C_1$, as shown in Fig.~\ref{fig:illustration}b, and another for measuring $C_2$, as depicted in Fig.~\ref{fig:circuit-corr-rhoFV-optimized}.

\subsection{Circuit transpilation and depth evaluation\label{section:appendix:depth-evaluation-corr}}
\begin{figure}[ht]
    \centering
    \includegraphics[width=0.4\textwidth]{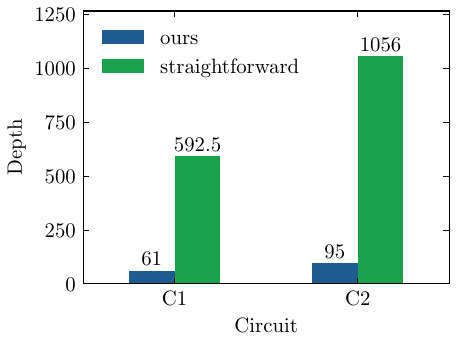}
    \caption[Depths of the transpiled circuits for $\mathcal{C}_{\mathcal{F}_{\rm corr}}$]{\label{fig:appendix:depth-corr}Median depths of the transpiled circuits for measuring $C_1$ and $C_2$ in $\mathcal{C}_{\mathcal{F}_{\rm corr}}$. The reported depths are based on 50 circuits, each corresponding to a different set of parameters.}
\end{figure}

Similar to the previous demonstration for $\mathcal{F}_{\rm opt}$, we conduct demonstrations on 50 sets of parameters described in Appendix~\ref{section:appendix:corr-system}. As shown in Fig.~\ref{fig:appendix:depth-corr}, we observe a significant reduction in the depths achieved by our implementations for both $C_1$ and $C_2$.

\subsection{Empirical results}\label{section:appendix:correlator-empirical-results}
\begin{figure*}[ht]
    \centering
    \includegraphics[width=\textwidth]{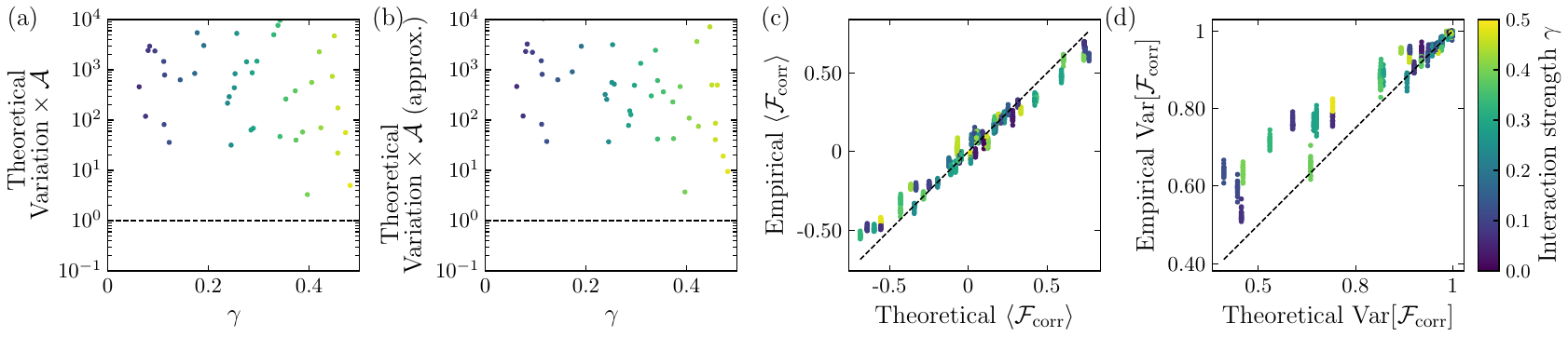}
    \caption[Theoretical, approximated, and empirical values regarding $\mathcal{F}_{\rm corr}$]{\label{fig:appendix:experiment-result-corr-Heron-var}Comparison of theoretical, approximated, and empirical values of $\mathcal{F}_{\rm corr}$. (a) Theoretical product of the variation ${\rm Var}[\mathcal{F}_{\rm corr}]/(\expval{\mathcal{F}_{\rm corr}}-\mathcal{C}_{\mathcal{F}_{\rm corr}})^2$ and survival activity $\mathcal{A}$ plotted against $\gamma$. (b) Theoretically approximated product of variation and $\mathcal{A}$. (c) Theoretical and empirical values of $\expval{\mathcal{F}_{\rm corr}}$. (d) Theoretical and empirical values of $\rm Var[\mathcal{F}_{\rm corr}]$. In (a) and (b), 50 data points are shown. In (c) and (d), the dashed line represents $y=x$, and $50\times 20$ data points are shown.}
\end{figure*}
\begin{figure*}[ht]
    \centering
    \includegraphics[width=\textwidth]{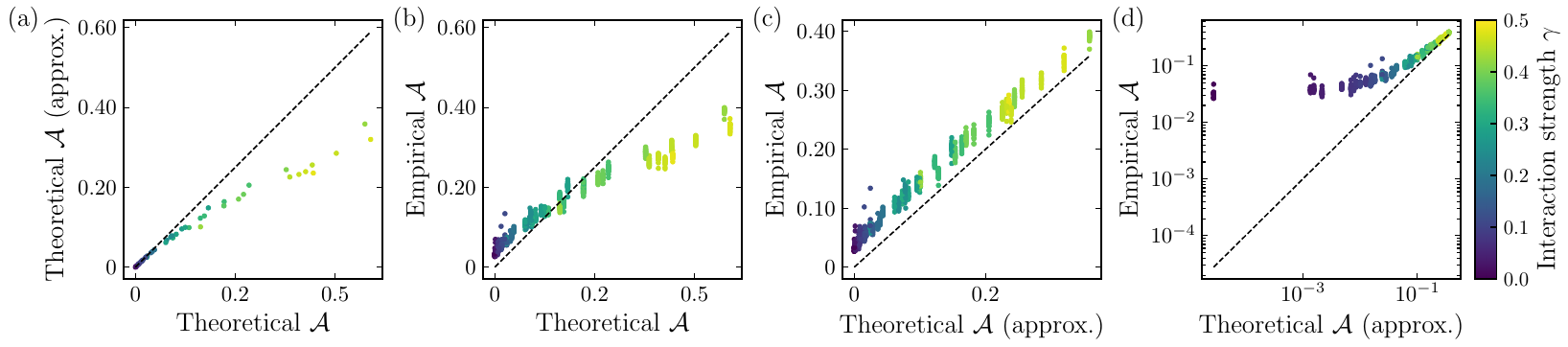}
    \caption[Theoretical, approximated, and empirical values of $\mathcal{A}$ for time correlators]{\label{fig:appendix:experiment-result-corr-Heron-A}Comparison of theoretical, approximated, and empirical values of the survival activity $\mathcal{A}$. (a) Theoretically approximated values plotted against theoretical values. (b) Empirical values plotted against theoretical values. (c) Empirical values plotted against theoretically approximated values. (d) Logarithmic plot of the data shown in (c). In panels (b)--(d), $50\times 20$ data points are displayed.}
\end{figure*}
\begin{figure*}[ht]
    \centering
    \includegraphics[width=\textwidth]{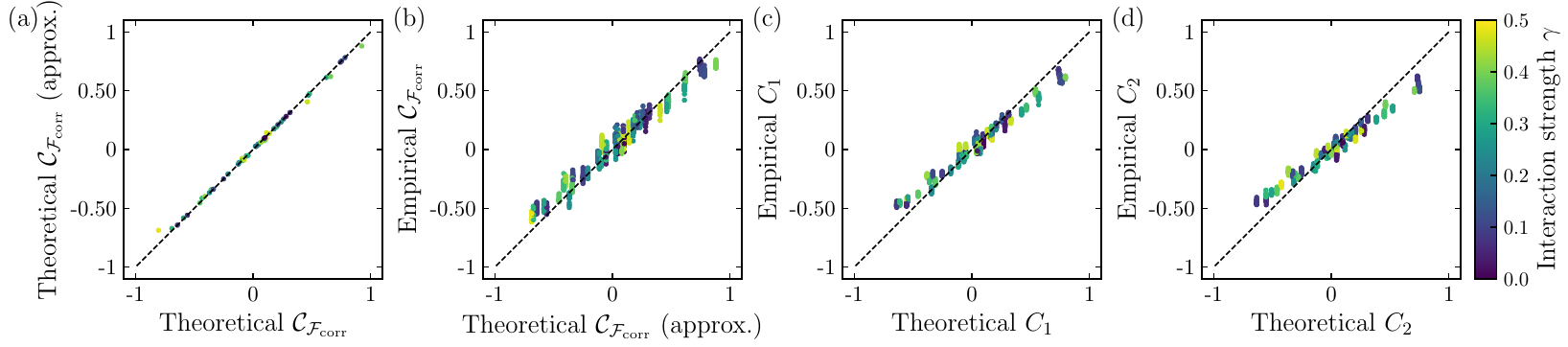}
    \caption[Theoretical, approximated, and empirical values of $\mathcal{C}_{\mathcal{F}_{\rm corr}}$]{\label{fig:appendix:experiment-result-corr-Heron-CF}Comparison of theoretical, approximated, and empirical values of $\mathcal{C}_{\mathcal{F}_{\rm corr}}$ and its component $C_1$ and $C_2$. (a) Theoretically approximated values of $\mathcal{C}_{\mathcal{F}_{\rm corr}}$ plotted against theoretical values. (b) Empirical values of $\mathcal{C}_{\mathcal{F}_{\rm corr}}$ plotted against theoretically approximated values. (c) Empirical values of $C_1$ plotted against theoretical values. (d) Empirical values of $C_2$ plotted against theoretical values. In panels (b)--(d), $50\times 20$ data points are displayed.}
\end{figure*}

We conducted demonstrations on 50 sets of parameters for $\theta_i$ and $\gamma$, as described earlier in Appendix~\ref{section:appendix:corr-system}. For each set, we performed measurements on the corresponding circuits 4,000 times. To evaluate the fluctuation of the results, we repeated this procedure 20 times.

In Fig.~\ref{fig:appendix:experiment-result-corr-Heron-var}, we show the influence of the theoretical approximation for $\mathcal{A}$ and $\mathcal{C}_{\mathcal{F}_{\rm corr}}$, as well as the empirical values of $\expval{\mathcal{F}_{\rm corr}}={\rm Re}[R(T)]$ and ${\rm Var}[\mathcal{F}_{\rm corr}]={\rm Var}[{\rm Re}[R(T)]]$. First, from Figs.~\ref{fig:appendix:experiment-result-corr-Heron-var}a and \ref{fig:appendix:experiment-result-corr-Heron-var}b, we observe that the influence of the theoretical approximation is not significant. We will later discuss the individual quantities $\mathcal{A}$ and $\mathcal{C}_{\mathcal{F}_{\rm corr}}$. From Fig.~\ref{fig:appendix:experiment-result-corr-Heron-var}c, we find that the absolute values of $\expval{\mathcal{F}_{\rm corr}}$ are smaller than the predicted values for relatively large $\gamma$, although the empirical values show good agreement with the predictions. Considering that $\expval{\mathcal{F}_{\rm corr}}={\rm Re}[R(T)]$, this trend is possibly due to unexpected noise that deteriorates the time correlators and that depolarizes the ancilla qubit. Finally, from Fig.~\ref{fig:appendix:experiment-result-corr-Heron-var}d, we observe that the empirical values of ${\rm Var}[\mathcal{F}_{\rm corr}]$ deviate from the predictions for small ${\rm Var}[\mathcal{F}_{\rm corr}]$. This is because ${\rm Var}[\mathcal{F}_{\rm corr}]$ is given by ${\rm Var}[\mathcal{F}_{\rm corr}]=1-\expval{\mathcal{F}_{\rm corr}}^2$, reflecting the deviation of $\expval{\mathcal{F}_{\rm corr}}$.

Next, we present the results of the survival activity $\mathcal{A}$ in Fig.~\ref{fig:appendix:experiment-result-corr-Heron-A}. From Fig.~\ref{fig:appendix:experiment-result-corr-Heron-A}a, we observe that the approximated values of $\mathcal{A}$ are consistent with the exact values for small $\gamma$, but smaller than the theoretical values for relatively large $\gamma$. This is because the approximation $\mathcal{A}\approx 1-p_0$ is the first-order approximation of the survival activity. Furthermore, from Fig.~\ref{fig:appendix:experiment-result-corr-Heron-A}b, we find that the empirical values of $\mathcal{A}$ exhibit the same trend as the approximated theoretical values shown in Fig.~\ref{fig:appendix:experiment-result-corr-Heron-A}a. Figure~\ref{fig:appendix:experiment-result-corr-Heron-A}c shows a comparison of the empirical and approximated values. We observe that although the empirical values generally align well with the predictions,  systematic deviations occur where the empirical values slightly exceed the predicted ones. This deviation is more clearly demonstrated in Fig.~\ref{fig:appendix:experiment-result-corr-Heron-A}d, where the empirical values do not decrease as the predictions suggest.

Figure~\ref{fig:appendix:experiment-result-corr-Heron-CF} displays the results of the inherent-dynamics contribution $\mathcal{C}_{\mathcal{F}_{\rm corr}}$. We find that the approximated values of $\mathcal{C}_{\mathcal{F}_{\rm corr}}$ align well with the exact values. Furthermore, from Fig.~\ref{fig:appendix:experiment-result-corr-Heron-CF}b, we find that the empirical values of $\mathcal{C}_{\mathcal{F}_{\rm corr}}$ are also consistent with the predictions. From Figs.~\ref{fig:appendix:experiment-result-corr-Heron-CF}c and \ref{fig:appendix:experiment-result-corr-Heron-CF}d, we observe that the absolute values of the empirical values of $C_1$ and $C_2$ are slightly smaller than the predicted values. Nevertheless, their empirical values align well with the predictions, highlighting the effectiveness of reducing the circuit depth. Note that the deviations of the empirical values of $C_2$ from the predictions are slightly larger than those of $C_1$, probably due to differences in circuit depth.

\subsection{Analysis of the empirical errors}\label{section:appendix:correaltor-error-analysis}
\begin{figure*}[ht]
    \centering
    \includegraphics[width=\textwidth]{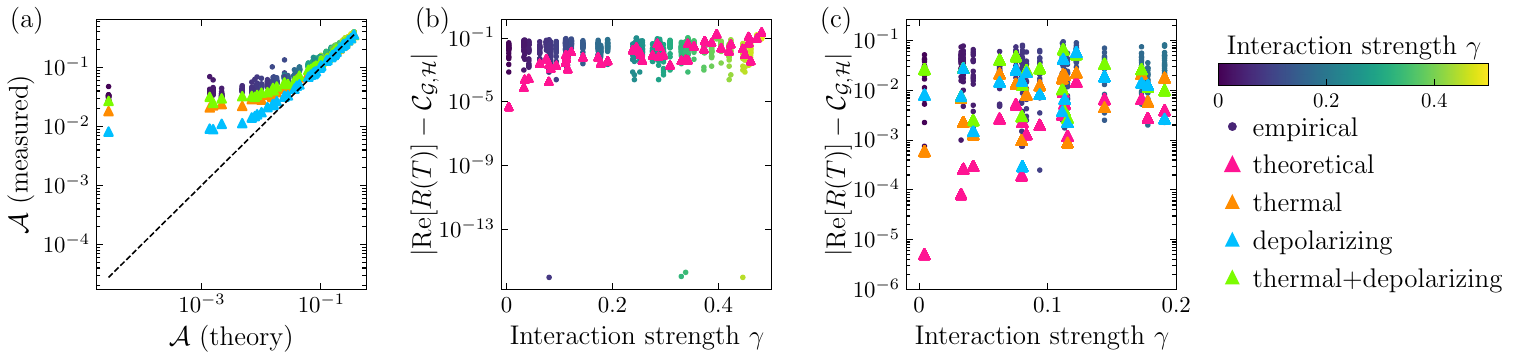}
    \caption[Noisy simulations of $\mathcal{A}$ for time correlators]{\label{fig:appendix:experiment-result-corr-Heron-thermal-A}Comparison between empirical results and noisy simulations for ${\rm Re}[R(T)]$. (a) Empirical and simulated values of $\mathcal{A}$ versus theoretical values. (b) Difference $|{\rm Re}[R(T)]-\mathcal{C}_{\mathcal{G},\mathcal{H}}|$ versus interaction strength $\gamma$. (c) Same as (b) but focused on the very weak coupling regime ($\gamma\leq0.2$), showing empirical data, theoretical predictions, and numerical simulations.}
\end{figure*}

\begin{figure*}[ht]
    \centering
    \includegraphics[width=\textwidth]{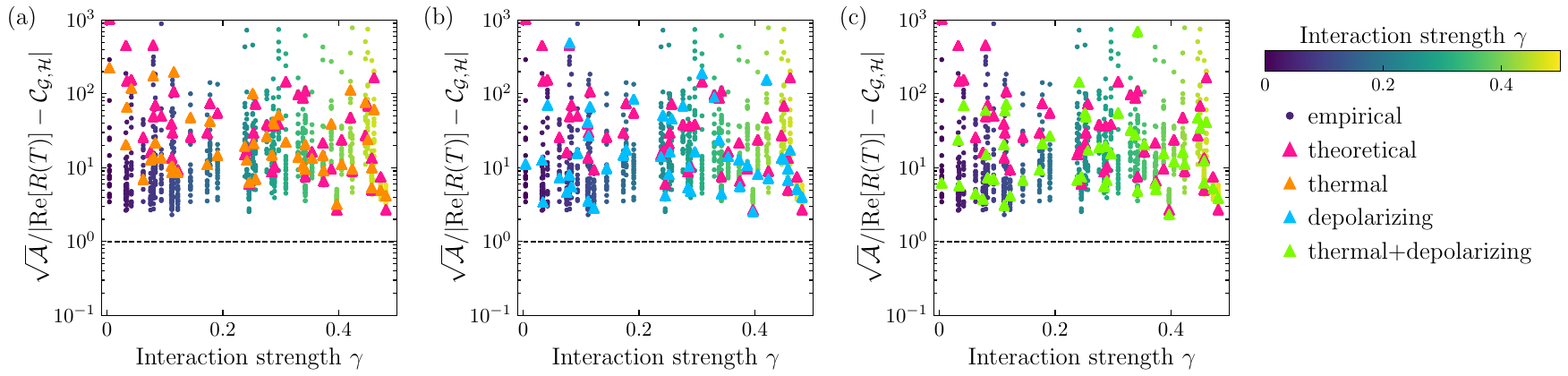}
    \caption[Nosy simulations of the inefficiency for time correlators]{\label{fig:appendix:experiment-result-corr-Heron-thermal-inefficiency}Comparison between empirical results and noisy simulations of the thermodynamic inefficiency $\mathcal{A}/|{\rm Re}[R(T)]-\mathcal{C}_{\mathcal{G},\mathcal{H}}|$. The simulations include: (a) thermal effects, (b) depolarizing errors, and (c) combined thermal effects and depolarizing errors.}
\end{figure*}

We analyze the source of the empirical errors through numerical simulations of physical circuits. Following our previous demonstration, we incorporate thermal initial states and depolarizing errors using Qiskit \cite{Qiskit_contributors_undated-lj}. Figure~\ref{fig:appendix:experiment-result-corr-Heron-thermal-A} shows the simulation results for both $\mathcal{A}$ and $|{\rm Re}[R(T)]-\mathcal{C}_{\mathcal{G},\mathcal{H}}|$. From Fig.~\ref{fig:appendix:experiment-result-corr-Heron-thermal-A}a, we find that thermal effects predominantly contribute to the empirical errors in $\mathcal{A}$. However, the situation differs for $|{\rm Re}[R(T)]-\mathcal{C}_{\mathcal{G},\mathcal{H}}|$. The comparison between empirical and theoretical values shown in Fig.~\ref{fig:appendix:experiment-result-corr-Heron-thermal-A}b reveals four anomalous data points with zero values. To highlight the essential features, we plot $|{\rm Re}[R(T)]-\mathcal{C}_{\mathcal{G},\mathcal{H}}|$ for $\gamma\leq0.2$ in Fig.~\ref{fig:appendix:experiment-result-corr-Heron-thermal-A}c alongside simulation results. We find that the empirical values significantly exceed theoretical predictions at small $\gamma$. Moreover, the theory-observation discrepancies are explained by depolarizing errors, rather than thermal effects. This contrasts with our findings for $\mathcal{A}$.

Based on these results, we analyze the deviation between empirical results and theoretical predictions of the thermodynamic inefficiency given by $\sqrt{\mathcal{A}}/|{\rm Re}[R(T)]-\mathcal{C}_{\mathcal{G},\mathcal{H}}|$. As shown in Fig.~2e and Fig.~\ref{fig:appendix:experiment-result-corr-Heron-thermal-inefficiency}a, the empirical values are lower than theoretically expected for small $\gamma$. We plot the results of numerical simulations with different noise sources in Fig.~\ref{fig:appendix:experiment-result-corr-Heron-thermal-inefficiency}, indicating that these discrepancies can be attributed to depolarizing errors. We note that similar deviations are observed for the general QTUR for ${\rm Re}[R(T)]$.

\end{document}